\documentclass[fleqn,usenatbib]{mnras}
\usepackage{newtxtext,newtxmath}
\usepackage[T1]{fontenc}
\usepackage{stfloats}
\usepackage{rotating}
\usepackage[utf8]{inputenc}     
\usepackage[english]{babel}     
\usepackage{booktabs,caption}
\usepackage{graphicx}        
\usepackage{enumitem}           
\usepackage{listings}           
\usepackage{threeparttable}
\usepackage{todonotes}          
\usepackage{longtable}
\usepackage{setspace}
\usepackage{relsize}
\usepackage{placeins}
\usepackage{pdflscape}
\usepackage{float}
\usepackage[export]{adjustbox}
\usepackage{amsfonts}
\usepackage{multicol}
\usepackage{blindtext}
\usepackage{booktabs,caption}

\DeclareRobustCommand{\VAN}[3]{#2}
\let\VANthebibliography\thebibliography
\def\thebibliography{\DeclareRobustCommand{\VAN}[3]{##3}\VANthebibliography}

\usepackage{graphicx}	
\usepackage{amsmath}	

\newcommand{\hi}{{\sc H\,i}\xspace}

\title[Abell 3408]{MeerKAT-derived \hi kinematics and the Baryonic Tully-Fisher Relation in the X-ray luminous cluster Abell 3408}

\author[Ndaliso et al.]{Xola Ndaliso,$^{1}$\thanks{E-mail: xola.ndaliso@gmail.com}
Tariq Blecher$^{4, 3}$,
Roger P. Deane$^{1,2}$,
Ed Elson$^{5}$
\\
$^1$Wits Centre for Astrophysics, School of Physics, University of the Witwatersrand, 1 Jan Smuts Avenue, Johannesburg, South Africa\\
$^2$ Department of Physics, University of Pretoria, Hatfield, Pretoria, 0028, South Africa\\
$^3$ South African Radio Astronomy Observatory, 2 Fir Street, Black River Park, Observatory, Cape Town, 7925, South Africa\\
$^4$ Centre for Radio Astronomy Techniques and Technologies, Department of Physics \& Electronics, Rhodes University, PO Box 94, Makhanda, 6140, South Africa\\
$^5$ Department of Physics and Astronomy, University of the Western Cape, Robert Sobukwe Rd, 7535 Bellville, Cape Town, South Africa
}

\date{Accepted 2025 August 17. Received 2025 July 18; in original form 2024 November 12}

\pubyear{2025}

\begin{document}
\label{firstpage}
\pagerange{\pageref{firstpage}--\pageref{lastpage}}
\maketitle

\begin{abstract}

\noindent  Significant advances in observational capabilities are continuously transforming our understanding of the dense environment of galaxy clusters and its impact on individual galaxies. Discerning between intrinsic and externally-induced properties of galaxies, including their gas kinematics, is a key diagnostic in the field of galaxy evolution. In this work, we present MeerKAT \hi spectral line observations of the redshift $\textit{z}$~$\sim$~0.042 galaxy cluster Abell 3408. A total of 64 galaxies are detected in \hi in this X-ray-luminous galaxy cluster (\textit{L}$_{X}$~$\sim$~3~$\times$~10$^{43}$~ergs s$^{-1}$). We model the \hi morphology and gas kinematics of the individual galaxies, using a semi-automated pipeline based on \textsc{CANNUBI} and \textsc{pyBBarolo}. The pipeline was developed and tested as part of this study. Of the 64 galaxies detected in the cluster, we successfully modelled 16, while the remaining galaxies exhibit disturbed \hi morphologies, insufficient angular or velocity resolution. We combine the galaxies with converged kinematic fits with 67 field galaxies from the MeerKAT spectral line survey early science data ($\langle$ \textit{z}~$\rangle$~=~0.0435) to produce a measurement of the Baryonic Tully-Fisher Relation (bTFr) that encompasses a broader range of environment and provides a useful comparison. We find a slope ($\alpha$~=~3.66$^{+0.32}_{-0.28}$) for this relation, which is consistent with that found from the MIGHTEE bTFr derived from the same definition. Interestingly, \hi detections of the Abell~3408 galaxy cluster galaxies are seen to extend the bTFr of the MIGHTEE sample, both in mass and velocity, despite their cluster environment. 

\end{abstract}



\section{Introduction}

In the hierarchical structure-formation model, the construction of significant structures like galaxy clusters occurs through the continuous process of merging and accretion of galactic haloes (see, e.g. \citealt{Mantz_2010}; \citealt{Wu_2010}; \citealt{White_2010}; \citealt{Tinker_2011}; \citealt{Kravstov_2012}; \citealt{Wei_2016}; \citealt{Planck_2016}; \citealt{Caminha_2022}). Galaxies fall into clusters in radial orbits (see, e.g. \citealt{Biviano_2013}; \citealt{Munari_2014}; \citealt{Mamon_2019}) and become susceptible to diverse astrophysical processes imposed by the hot, ionized intra-cluster medium (ICM) of the dense environment. These astrophysical mechanisms impact the equilibrium, morphologies, gas content, and star-formation activity of the galaxies as they traverse the ICM (\citealt{Serra_2017}).\\

The gas content of cluster galaxies is primarily influenced by environmental processes, including hydrodynamical mechanisms such as ram-pressure stripping, where cold gas is removed from the interstellar-medium (ISM) by the hot intracluster-medium (ICM) (e.g. \citealt{Gunn_Gott_1972}; \citealt{vulcani_2023}; \citealt{Piraino_2023}; \citealt{obeirne_2024}) — and starvation (or strangulation), a process in which gas accretion onto the galaxy is halted after infall into a cluster halo, gradually quenching star formation over long timescales (e.g. \citealt{Cortese_2021}; \citealt{Boselli_2022}. The former mechanism halts the star-formation activity in the galaxy disc on longer time scales ($\sim$4 Gyr; \citealt{Boselli_2014}), as compared to the latter, which ceases star-formation on relatively short time scales ($\sim$100 - 200 Myr; \citealt{Vollmer_2004}), although recent studies suggest that in some cases, ram-pressure-induced quenching may take place over longer timescales of 1.5–3.5 Gyr, comparable to starvation (\citealt{Marasco_2023}). Ram-pressure stripping, amongst other processes, has been reported to be the principal mechanism that reduces the star-formation rates in cluster galaxies (e.g. \citealt{Koopmann_2004}). The loss of gas in cluster galaxies has been explained further with other processes that occur in galaxy clusters such as turbulent viscous stripping (e.g. \citealt{Chung_2007}), and thermal evaporation (e.g. \citealt{Nipoti_2010}). The cold gas affected by these mechanisms includes neutral atomic hydrogen (\hi) and the molecular gas, which can also experience environmental effects (e.g., \citealt{Lee_2016}; \citealt{Brown_2021}; \citealt{Boselli_2022}; \citealt{Zabel_2022}). However, molecular gas and dust are generally more resilient compared to diffuse phases such as \hi and ionized gas (e.g., H$\alpha$-emitting gas, \citealt{Poggianti_2017}).
\\

Over the past 60 years, studies of galaxy rotation curves have played a pivotal role in understanding the matter distribution in galaxies, revealing a discrepancy between the observed velocities of stars and gas and those expected based solely on the visible matter, thereby providing the cornerstone of evidence for dark matter (e.g. \citealt{Rubin_1978}; \citealt{Bosma_1981}; \citealt{vanAlbada_1985}). The distribution of dark matter in galaxies has been an open question and a long-standing one, for several decades, the 'cusp/core' problem as it is commonly called (\citealt{de_Blok2008}). \\

Galaxy rotation curves have taken a spot-light, in attempts to solve this discrepancy. The neutral hydrogen (\hi), often extends to several optical diameters, tracing galaxy dynamics at large radii and into regions where the dark matter is thought to dominate the dynamics of galaxies ( \citealt{de_Blok_2002}; \citealt{de_Blok2008}; \citealt{Swaters_2012}; \citealt{Lelli_2016}; \citealt{Pavel_2021}; \citealt{laudage2024}). \hi follows nearly circular orbits in galaxies and for its diffuse and extended nature, serves as one of the leading tracers for studying rotation curves of galaxies. However, historically, \hi rotation curves have been limited to small samples of galaxies. It is only over the past 2 decades we have seen an increase in high-resolution \hi surveys of galaxies, such as: The \hi Nearby Galaxy Survey (THINGS, \citealt{Walter_2008}), Very Large Array survey of Advanced Camera for Surveys Nearby Galaxy Survey Treasury galaxies (VLA-ANGST, \citealt{Ott_2012}), Local Irregulars That Trace Luminosity Extremes THINGS (LITTLE THINGS, \citealt{Hunter_2012}), Local Volume \hi Survey (LVHIS, \citealt{Koribalski_2018}), enabling high-resolution rotation curves of local ($\lesssim$ 12 Mpc) galaxies. These surveys have enhanced the resolution of \hi rotation curves, enabling the construction of high-quality mass models. For instance, \cite{Mancera_Pina_2022} developed resolved kinematic and mass models for 32 LITTLE THINGS galaxies, which form a subsample of the study conducted by \cite{Iorio_2017} and a subset of nearby spiral galaxies analyzed by \cite{Di_Teodoro_2021}.\\

Larger samples of high-quality, highly resolved \hi rotation curves have, in turn significantly advanced the study of the dynamical scaling relations such as the Tully-Fisher relation (TFr; \citealt{Tully_1997}). The TFr is a dynamical relation that links a spiral galaxy's mass-luminosity to its rotation velocity. This relation has proven to be important in understanding galaxy formation and evolution, and several authors (e.g. \citealt{Verheijen_2001}; \citealt{Lelli_2016}; \citealt{Ponomareva_2017} ) have studied the TFr across a broad class of galaxies. \cite{Chung_2002} studied and showed that TFr holds rotating galaxies of different morphologies. Similarly \cite{Melgarejo_2021}, using optical observations, showed that TFr holds for galaxies in dense environments at $\textit{z}$ $\sim$ 0.7. \cite{Courteau_2003} also studied the TFr for un/barred spiral galaxies, finding no correlation between the presence of a bar in a galaxy and its projection on the TF plane. This provided further insights that un/barred spirals have, on average, comparable fractions of luminous- and dark-matter fractions at a given radius.\\

At $\textit{z}$ $\sim$ 0, \cite{Lelli_2019} studied the baryonic TFr (bTFr; \textcolor{red}{\citealt{McGaugh_2000}}) relation for 153 galaxies in the $\textit{SPARC}$ (Spitzer Photometry $\&$ Accurate Rotation Curves) database, consisting of 175 spiral galaxies with high-quality \hi rotation curves (\citealt{Lelli_2016}). This is considered the largest observational study of the bTFr to date. In their work, they studied the bTFr based on different velocity definitions such as $\textit{W}_{50}$, which is the rotational velocity measure derived from the integrated \hi line profile; and $\textit{V}_{\rm flat}$, which is the velocity measured on the flat part of the \hi rotation curve and $\textit{V}_{\rm max}$ - the maximum measured rotational velocity from the \hi rotation curve. The tightest bTFr with the steepest slope was achieved using the $\textit{V}_{\rm flat}$ velocity definition. These results are in good agreement with recent studies of the bTFr based on \textsc{SIMBA} (see \citealt{Dave_2019}) cosmological hydrodynamical simulations (\citealt{Glowacki_2020}). \cite{Ponomareva_2021} used the MIGHTEE (see \citealt{jarvis_2017} for details on the survey) Early Science data release to perform, for the first time, a homogeneous study of the \hi-based bTFr over the last billion years (0 $\lesssim$ \textit{z} $\leq$ 0.081) using a sample of 67 field galaxies. They considered two velocity measures: the $\textit{W}_{50}$ from the corrected width of the \hi global profile, and $\textit{V}_{\rm out}$, the rotational velocity measured at the outermost point of the resolved \hi rotation curves. For both velocity definitions, they found relations with very low intrinsic scatter orthogonal to the best-fitting relation, comparable to the $\textit{SPARC}$ sample at $\textit{z}$ $\sim$ 0. They also found that the slopes of the relations are consistent with several studies carried at $\textit{z}$~$\sim$~0.\\

In our work, we use MeerKAT \hi spectral line observations of the redshift \textit{z} $\sim$ 0.042 galaxy cluster Abell 3408. For the first time, we model the kinematics of detected Abell 3408 cluster galaxies in a uniform way using a pipeline, developed as part of this work, based on \textsc{CANNUBI} and \textsc{pyBBarolo}. We successfully model the kinematics of 16 galaxies within this cluster. Furthermore, we use the converged kinematics  models to study the \hi-based bTFr, extending our sample with 67 MeerKAT MIGHTEE field galaxies from \cite{Ponomareva_2021}.\\

The structure of the paper is as follows. In Section~\ref{sec:2} we describe the MeerKAT observations, and data reduction and present the \hi data products of the Abell 3408 galaxy cluster in Section~\ref{sec:3}. In Section~\ref{sec:4} we discuss the \hi kinematic modelling pipeline developed as part of this work, we also present the rotation curves overlaid on position-velocity slices of the 16 modelled galaxies. The \hi-based bTFr and the fitting procedure are discussed in Section~\ref{sec:5} with summary and conclusions in Section~\ref{sec:6}. Throughout this work, we assume cosmological values $\Omega_{\rm M}$~=~0.311, $\Omega_{\rm \Lambda}$~=~0.688, and \textit{H}$_{\rm 0}$~=~67.66~km\,s$^{-1}$ (\citealt{Planck_2018}).

\section{MeerKAT Observations and Data Processing}
\label{sec:2}

Abell 3408 was observed with MeerKAT (\citealt{Jonas2016TheMR}~$\&$~MeerKAT team) on 11, 12 of May 2019 with 61 antennas. The data were taken using the \textit{L}-band receivers (900 - 1670 MHz), with a bandwidth of 856 MHz divided into 4096 channels (see Table \ref{tab:obs_summary}). The 209 kHz channel width has an equivalent velocity width of $\sim$ 46 km\ s$^{\rm -1}$ at \textit{z} $\sim$ 0.042. The full observation duration was $\sim$ 11.6 hrs with integration time set to eight seconds. J0408-6545 was used as a primary flux and bandpass calibrator, and was observed for 10 minutes after every 2-hour scan of the target. The gain calibrator, J0825-5010 was observed for 2 minutes after every 1 minute scan of the target. \\ 

\subsection{Calibration and Imaging}
\label{subsec:2.1}

We process a bandwidth subset of 100 MHz, from 1300-1400 MHz, which encompasses all the individual cluster members as well as the candidate lensed \hi source. The radio frequency interference (RFI) was flagged using the \textsc{AOFlagger} (\citealt{Offringa_2010}). We used an early version of the \textsc{oxkat} data processing software for calibration (see \citealt{Heywood_2020}, for full details). The antenna-based terms were solved for, in the following order: geometric delays, initial antenna gain phase, complex bandpass response, complex antenna gains (with the initial antenna gain phase discarded), and absolute flux scaling.\\

In the self-calibration step, we adopted an iterative approach between imaging, setting the \texttt{Briggs robust} parameter to -0.3, and phase-only self-calibration using a solution interval of 20 seconds. Overall flagging percentages of $\sim$ 10 per cent in both XX and YY polarizations, across the 100 MHz bandwidth, were measured after calibration. This percentage is significantly lower than the typical full band amounts, this is because the 100 MHz is in a generally clean region within the \textit{L}-band.\\ 

The continuum emission across the 100 MHz bandwidth was imaged and modelled using \textsc{wsclean} (\citealt{Offringa_2010}) to create a multi-frequency synthesis (MFS) image with four 25 MHz sub-bands. A multi-stage approach was adopted for the deconvolution step. Initially, an MFS continuum image was created using auto-masking with the masking threshold set to 10 $\sigma_{\rm global}$, where $\sigma_{\rm global}$ is the global RMS of the entire image. Secondly, using the \textsc{ddfacet} package (\citealt{Tasse_2018}) a static mask was derived from this image. To avoid artefacts with bright sources, a local RMS threshold of 6 $\sigma_{\rm local}$ was used (\citealt{Heywood_2020}). Thirdly, a new MFS continuum image was created using the static mask as well as the auto-masking with a similar setting as before. Lastly, the image was manually inspected, and we found that the procedure successfully deconvolved both low and high-brightness sources.\\

To remove the continuum emission, we subtracted a frequency-linear continuum visibility model from the calibrated visibilities.  A \textsc{clean} cube is then generated with \textsc{wsclean} using an auto-masking threshold of 5$\sigma_{\rm global}$. To mitigate the residual continuum in the emission cube, we then performed image-plane continuum subtraction by fitting a frequency-dependent first-order polynomial to each pixel with the \textsc{casa} task \textsc{imcontsub}. After this step, the continuum artefacts were largely within the thermal noise level. With the resulting cube, we reached a 209 kHz channel noise of 85~$\mu$Jy beam$^{\rm -1}$ using \texttt{Briggs robust} of 0.5 weighting.\\

To correct the effect of the primary beam, we generated a MeerKAT primary beam model with the \textsc{eidos} package (\citealt{Asad_2021}). The parameters were set to produce a Stokes I beam model with the same frequency and angular dimensions as the data cube. The data cube was then corrected by dividing the intensity of each pixel, with the size of 3$^{\prime\prime}$, by the Stokes I beam model at that position. The resulting PSF FWHM is 13.1$^{\prime\prime}$ $\times$ 11.7$^{\prime\prime}$ and the position angle is -42.6 deg East of North. 

\subsection{Source finding and Cross-matching}
\label{subsec:2.2}

The Source Finding Application (\textsc{sofia}, \citealt{Serra_2015}) was run on the final primary-corrected beam cube, to extract the spectral line detections and source properties. \textsc{sofia} offers several functions to improve the reliability of detections. The procedure used in this work is as follows:

\begin{enumerate}[label=\roman*., leftmargin=1cm]
    \item smoothing the cube iteratively and extracting only pixels above a certain threshold,
    \item merging the pixels to a detection threshold
    \item reliability filtering
    \item mask dilation
    \item source parametrisation
\end{enumerate}

\begin{table}
	\centering
	\caption{Summary of \textit{L}-band MeerKAT observations of Abell 3408. }
	\label{tab:obs_summary}
	\begin{tabular}{lc} 
		\hline
		\hline\\
		Observing dates & 11/05/2019-12/05/2019\\
		Pointing centre (J2000) & 07h08m31.7s -49d12m52s\\
		No. of antennas & 61\\
		On-source time  & 11.6 (hrs)\\
		t$_{\rm int}$  & 8 (sec)\\
		Frequency range  & 856-1712 (MHz)\\
		$\rm \Delta \nu$ & 209 (kHz)\\
		    & 46 km s$^{\rm -1}$ for \hi at z = 0.04\\
		\hi cube weighting & \texttt{Briggs robust} 0.5\\
		$\rm \sigma_{rms}$  per 209 kHz channel & 85 ($\rm \mu$Jy beam$^{\rm -1}$)\\
		$\rm \theta_{synth}$, BPA  & 13.1 $\times$ 11.7, -42.6 ($^{\prime \prime}$ ,$\times$ $^{\prime \prime}$, $^{\circ}$)\\
		\hline
	\end{tabular}
\end{table}

The cube was smoothed using a boxcar convolution with angular and frequency sizes up to (30$^{\prime\prime}$, 209 kHz). To ensure that not only bright pixels were extracted and to enable reliability calculation, the detection threshold was set to 3.8$\rm \sigma$ following iterative manual testing. Pixels with flux greater than this threshold were merged to the same source if separated by less than 18$^{\prime\prime}$ and 90 km s$^{\rm -1}$. False positives were filtered out by using the built-in reliability filter, with a reliability threshold of 90 per cent and a signal-to-noise ratio cut of 15 (see \citealt{Serra_2015} for more details).\\

Following this procedure, a total of 64 \hi detections, presented in Figure \ref{fig:moment_maps}, were found in the in the Abell~3408 galaxy cluster. We then cross-matched the \hi detections with the four mid-infrared bands (3.4, 4.6, 12, and 22 $\rm \mu m$) of the WISE all-sky survey (\citealt{Wright_2010}). As the \hi distribution can be off-centre its host galaxy due to environmental factors, we cross-matched the catalogue manually by overlaying the \hi intensity maps on the W1 and DSS \textit{R}-band maps (Figure \ref{fig:id_9_moment_maps}).\\

Of the 64 MeerKAT \hi detections in Abell 3408, we find counterparts for 55 sources ($\sim$ 86 per cent) in the 3.4 $\rm \mu m$ (W1 band), 50 sources ($\sim$ 78 per cent) in the 4.6 $\rm \mu m$ (W2 band), 41 sources ($\sim$ 64 per cent) in the 12 $\rm \mu m$ (W3 band), and 25 sources ($\sim$ 39 per cent) in the 22 $\rm \mu m$ (W4 band). \\

\section{\hi properties of the meerkat detections}
\label{sec:3}

\begin{figure*}
    \centering
    \includegraphics[scale=.8]{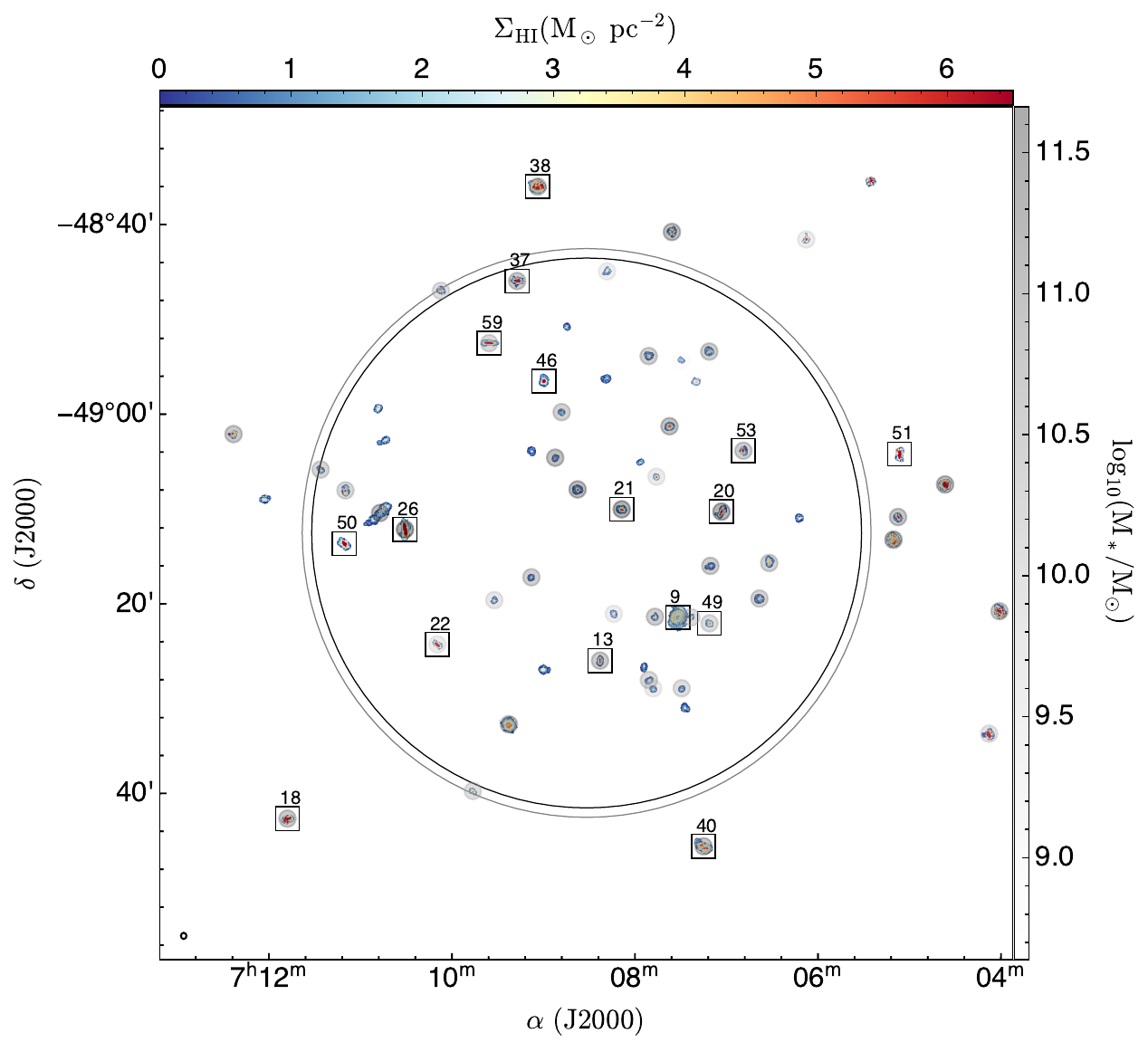}
    \caption{The MeerKAT \hi moment 0 map of the Abell 3408 galaxy cluster showing all the 64 detections coloured by the \hi mass surface density. Galaxies with W1 (3.4 $\mu$m) and W2 (4.2 $\mu$m) counterparts are shown with the grey shaded circles, coloured with \textit{M}$_{*}$ of each galaxy. The black squares mark the 16 galaxies with converged kinematic fits. The MeerKAT synthesized beam (13.1 $\times$ 11.7 arcsec) is shown in the bottom left corner. The black circle presents the MeerKAT Stokes I primary beam half maximum sensitivity and the grey circle marks the cluster radius (\textit{R}$_{200}$ = 1.5 Mpc).}
    \label{fig:moment_maps}
\end{figure*}

The MeerKAT \hi zeroth-order moment map of the Abell 3408 galaxy cluster showing all the 64 detections is presented in Figure \ref{fig:moment_maps}. This moment map is generated by \textsc{sofia} across the masked \hi data cube. \\

\begin{table*}
    \begin{center}
        \begin{threeparttable}
            \tabcolsep=3.5pt\relax
            \renewcommand{\arraystretch}{1.1}
            \caption{Properties of the 16 galaxies with converged kinematic fits.}
            \label{tab:properties_table}
            \begin{tabular}{lccccccccccr}
                \hline
            ID & $\rm \alpha\ (J2000)$ & $\rm \delta (J2000)$ & log(\textit{M}$\rm_{HI}$) & log(\textit{M}$\rm_{dyn}$) & log(\textit{M}$\rm_{*}$)  & \textit{w}$\rm_{50}$& \textit{z}$\rm_{ HI}$ & \textit{d}$\rm_{HI}$ & \textit{f}$\rm ^{*}_{HI}$ & \textit{f}$\rm^{ dyn}_{HI}$ \\
            -- & h:m:s & d:m:s & M$_{\odot}$ & M$_{\odot}$ & M$_{\odot}$ & km\ s$^{-1}$ & -- & kpc & -- & --  \\
            \small (1) & (2) & (3) & (4) & (5) & (6) & (7) & (8) & (9) & (10) & (11) \\
                \hline
                41 & 7 04 40.584  & -49 07 0.583  & 10.2 $\pm$  0.01 & 11.3 $\pm$ 0.06 & 11.2 $\pm$  0.12 & 95.4 $\pm$  20.5 & 0.043 & 72.1 $\pm$  9.4 & 0.91 $\pm$ 0.02 & 0.91 $\pm$  0.12 \\
                53 & 7 06 50.738  & -49 04 0.544  & 9.6 $\pm$  0.05 & 10.9 $\pm$ 0.05 & 10.6 $\pm$  0.12 & 328.3 $\pm$  34.8 & 0.043 & 36.2 $\pm$  7.0 & 0.9 $\pm$ 0.11 & 0.88 $\pm$  0.14 \\
                20 & 7 07 4.704  & -49 10 4.903  & 9.8 $\pm$  0.02 & 11.2 $\pm$ 0.05 & 11.4 $\pm$  0.12 & 580.0 $\pm$  59.0 & 0.044 & 51.7 $\pm$  7.9 & 0.86 $\pm$ 0.04 & 0.88 $\pm$  0.11 \\
                49 & 7 07 12.086  & -49 22 2.018  & 9.5 $\pm$  0.01 & 11.2 $\pm$ 0.05 & 10.2 $\pm$  0.12 & 128.4 $\pm$  19.8 & 0.043 & 36.2 $\pm$  7.0 & 0.93 $\pm$ 0.02 & 0.85 $\pm$  0.1 \\
                40 & 7 07 15.322  & -49 45 3.355  & 10.2 $\pm$  0.02 & 11.3 $\pm$ 0.05 & 11.0 $\pm$  0.12 & 323.3 $\pm$  34.3 & 0.047 & 82.2 $\pm$  10.2 & 0.93 $\pm$ 0.04 & 0.9 $\pm$  0.11 \\
                50 & 7 07 23.957  & -49 21 4.881  & 10.1 $\pm$  0.03 & 11.6 $\pm$ 0.05 & --- & 292.9 $\pm$  31.5 & 0.044 & 61.9 $\pm$  8.6 & --- & 0.87 $\pm$  0.12 \\
                9 & 7 07 32.609  & -49 21 4.913  & 10.4 $\pm$  0.02 & 10.9 $\pm$ 0.05 & 11.1 $\pm$  0.12 & 361.9 $\pm$  37.9 & 0.045 & 112.6 $\pm$  12.8 & 0.93 $\pm$ 0.04 & 0.95 $\pm$  0.11 \\
                21 & 7 08 9.061  & -49 10 2.029  & 9.7 $\pm$  0.03 & 10.7 $\pm$ 0.05 & 11.5 $\pm$  0.12 & 338.3 $\pm$  35.7 & 0.044 & 41.4 $\pm$  7.3 & 0.84 $\pm$ 0.05 & 0.91 $\pm$  0.11 \\
                13 & 7 08 23.014  & -49 26 9.881  & 9.6 $\pm$  0.04 & 10.3 $\pm$ 0.07 & 10.7 $\pm$  0.12 & 375.6 $\pm$  39.2 & 0.045 & 41.4 $\pm$  7.3 & 0.9 $\pm$ 0.07 & 0.93 $\pm$  0.16 \\
                46 & 7 08 23.014  & -49 26 9.881  & 9.9 $\pm$  0.02 & 11.7 $\pm$ 0.06 & --- & 188.2 $\pm$  22.8 & 0.043 & 46.6 $\pm$  7.6 & --- & 0.84 $\pm$  0.11 \\
                38 & 7 09 2.971  & -48 36 9.699  & 10.4 $\pm$  0.02 & 11.1 $\pm$ 0.05 & 10.9 $\pm$  0.12 & 229.1 $\pm$  26.0 & 0.044 & 92.4 $\pm$  11.0 & 0.96 $\pm$ 0.04 & 0.94 $\pm$  0.11 \\
                37 & 7 09 16.174  & -48 46 7.1000  & 9.9 $\pm$  0.02 & 11.2 $\pm$ 0.05 & 10.8 $\pm$  0.12 & 338.0 $\pm$  35.7 & 0.042 & 51.7 $\pm$  7.9 & 0.92 $\pm$ 0.04 & 0.89 $\pm$  0.11 \\
                59 & 7 09 34.243  & -48 52 9.544  & 10.0 $\pm$  0.01 & 11.3 $\pm$ 0.05 & 10.3 $\pm$  0.12 & 292.5 $\pm$  31.5 & 0.041 & 41.4 $\pm$  7.3 & 0.98 $\pm$ 0.02 & 0.89 $\pm$  0.10 \\
                22 & 7 10 8.491  & -49 24 4.344  & 9.9 $\pm$  0.03 & 10.5 $\pm$ 0.06 & 9.7 $\pm$  0.12 & 203.7 $\pm$  24.0 & 0.044 & 51.7 $\pm$  7.9 & 1.02 $\pm$ 0.07 & 0.94 $\pm$  0.14 \\
                26 & 7 10 29.009  & -49 12 6.182  & 10.4 $\pm$  0.02 & 10.6 $\pm$ 0.05 & 11.7 $\pm$  0.12 & 422.9 $\pm$  43.7 & 0.044 & 82.2 $\pm$  10.2 & 0.89 $\pm$ 0.04 & 0.98 $\pm$  0.12 \\
                18 & 7 11 46.922  & -49 42 9.579  & 10.2 $\pm$  0.01 & 11.4 $\pm$ 0.05 & 10.7 $\pm$  0.12 & 276.4 $\pm$  30.1 & 0.044 & 61.9 $\pm$  8.6 & 0.95 $\pm$ 0.03 & 0.9 $\pm$  0.10 \\
                \hline
		\end{tabular}
            \begin{tablenotes}
                \small
                \item[] \textit{Notes:} Cols (2)-(3): RA and DEC (J2000) coordinates of the detections from \textsc{sofia}. Cols (4)-(6): $\log(\textit{M}_{\rm HI})$ calculated from the \hi moment maps of the detections, $\log(\textit{M}_{\rm dyn})$ estimated from the last measured point of the \hi rotation curves, and $\log(\textit{M}_{\rm *})$ estimated from WISE. Cols (7)-(9): \textit{w}$_{50}$ from the \textsc{sofia}, \hi redshifts, and \hi diameters measures at 1 M$_{\odot}\,\rm pc^{-2}$. Cols (10)-(11): \hi to stellar, and dynamical mass fractions.
            \end{tablenotes}
        \end{threeparttable}
    \end{center}
\end{table*}

We observe that the majority of sources are positioned in the northwestern region of the galaxy cluster. This distribution may be influenced by the neighbouring galaxy cluster, Abell 3407, which was reported to interact with Abell 3408 (\citealt{Nascimento_2016}). The MeerKAT observations increase the number of known cluster members in the Abell 3408 galaxy cluster from the 27 galaxies previously reported in \cite{Nascimento_2016}.\\

\begin{figure*}
    \centering
    \includegraphics[scale=.6]{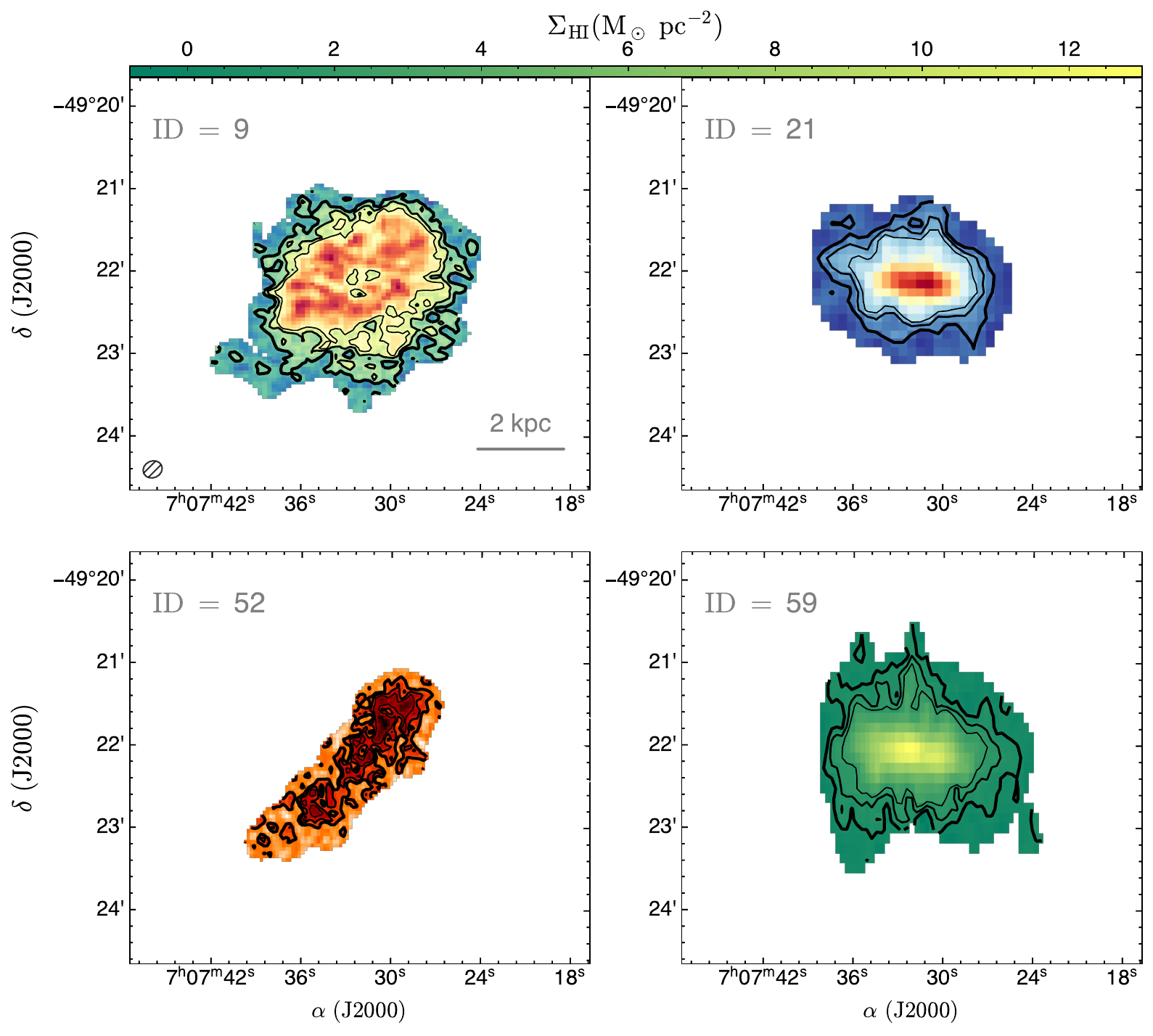}
    \caption{MeerKAT \hi moment zero maps of four sources that are representative of the 64 detected galaxies. The contour levels, with decreasing line-width, trace \hi surface densities at 1, 2, and 3 $\sigma_{\rm 3T}$ M$_{\odot}$\,pc$^{-2}$. The MeerKAT synthesized beam (13.1 $\times$ 11.7 arcsec) is displayed in the top left panel, with the grey horizontal bar indicating the corresponding physical scale.}
    \label{fig:four_maps}
\end{figure*}

We present MeerKAT \hi moment-zero maps of four detections that are representative of the \hi morphologies in the cluster in physical units of \hi surface densities i.e. M$_{\odot}$\,pc$^{-2}$. To convert the map from the flux units of Jy\,beam$^{-1}$\,Hz to units of Jy\,km\,s$^{-1}$ we divide the moment-zero map by the channel width ($\rm \Delta \nu$) in Hz, given in Table~\ref{tab:obs_summary} and multiply by the channel width in km\,s$^{-1}$, which results in a moment-zero map in units of Jy\,beam$^{-1}$\,km\,s$^{-1}$. To get the map in Jy\,km\,s$^{-1}$ we then divide by the beam area over the pixel area (see Eqn. 3 of \citealt{Iorio_2017}). This results in a map of units, Jy\,km\,s$^{-1}$, which can be converted to units of M$_{\odot}$ using Eqn. \ref{eqn:hi_mass}. This is then divided by the physical area of the pixel in pc$^2$, resulting in a moment-zero map with units of M$_{\odot}$\,pc$^{-2}$. Of these four detections shown in Figure~\ref{fig:four_maps}, the first detection (i.e. top left) is identified with ID 9 in the MeerKAT \hi cluster moment-zero map shown in Figure \ref{fig:moment_maps}, found with three other potential companions in close proximity, with a stellar mass of $\log_{10}$(\textit{M}$_{*}$)~=~11.13~M$_{\odot}$. The MeerKAT \hi moment-zero map of this source shows a significant drop in \hi surface densities in the central part of the disc. For the \hi moment maps, we follow the method outlined by \citet{Verheijen_Sancisi_2001} and \citet{Lelli_2014} to determine the outermost \hi contour level. In line with the approach of \citet{Iorio_2017}, we construct a signal-to-noise ratio (S/N) map for each galaxy and compute the mean value of the pixels with S/N between 2.75 and 3.25. This mean is then adopted as the 3$\sigma$ pseudo-level, denoted as $\sigma_{\rm 3T}$. \\

The second detection shown in Figure~\ref{fig:four_maps}, identifies with ID 21 in Figure \ref{fig:moment_maps}, which is seen almost at the cluster's centre, at a projected distance of \textit{d}$_{\rm proj}$~=~0.24 Mpc, with no immediate neighbouring detections, and a stellar mass of $\log_{10}$(\textit{M}$_{*}$) = 11.48 M$_{\odot}$. The majority of the \hi mass is concentrated in the centre of this galaxy, with contours showing signs of marginal disturbance on the \hi disc, which might be induced by the intra-cluster medium. This is seen from the second contour at 3 M$_{\odot}$\,pc$^{-2}$ which is asymmetric along the minor axis of this source. The third moment-zero map is an example of a galaxy undergoing ram-pressure stripping. This detection is ID'ed 52 in Figure \ref{fig:moment_maps}, with a stellar mass of $\log_{10}$(\textit{M}$_{*}$)~=~11.17 M$_{\odot}$, and it seems to be undergoing or have undergone ram-pressure stripping. The \hi gas in this galaxy appears to be stripped by the intra-cluster medium, forming a long \hi tail. Additionally, at least based on the \hi moment 0 map of this galaxy, gas accretion via merger could also be driving this extended \hi tail. The last moment-zero map is that of detection ID 59, which is also seen quite isolated from Figure \ref{fig:moment_maps}. The inner \hi disc of this source does not show any signs of disturbances, i.e. it is regular. However, well outside the stellar radius, the contours show that this galaxy might be experiencing some astrophysical mechanisms imposed by the intra-cluster medium. The rest of the MeerKAT \hi moment-zero maps for all the 64 galaxies are presented in Appendix \ref{sec:appendix_a}.\\

\begin{figure*}
    \centering
    \includegraphics[scale=.58]{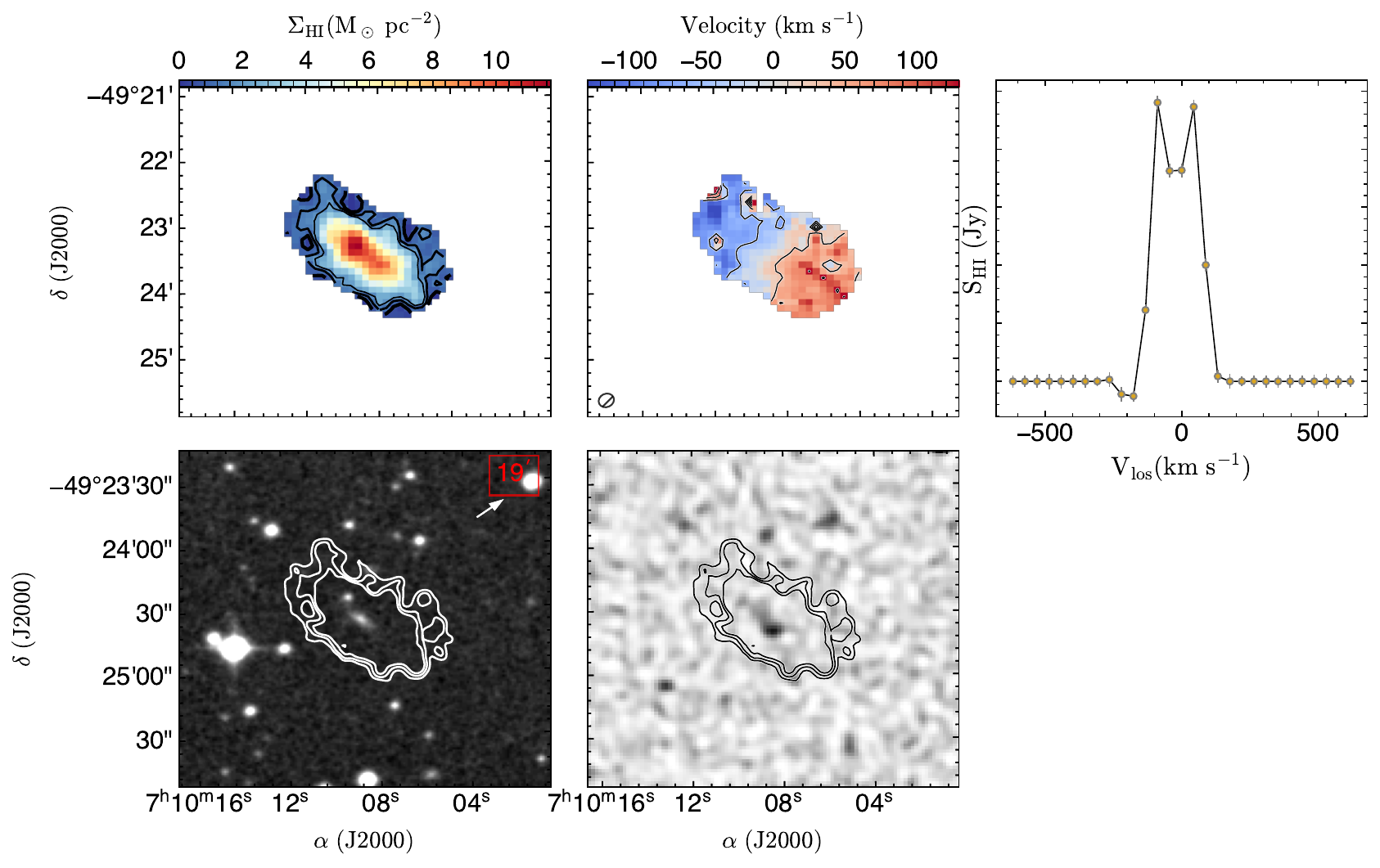}
    \caption{\textbf{Top Left}: The MeerKAT \hi intensity map of the galaxy with source ID 22 is presented with contour levels at 1, 2, and 3 $\sigma_{\rm 3T}$ M$_{\odot}$\,pc$^{-2}$.  \textbf{Top Mid}: The \hi velocity field of the same galaxy. \textbf{Top Right}: The integrated \hi spectrum in the upper right corner, with uncertainties calculated as RMS of each channel in the \hi data cube. \textbf{Bottom Left}: The DSS \textit{R}-band image with \hi contours overlaid. The distance in arcminutes and direction to the cluster centre are shown by the red text in the box and the white arrow, respectively. \textbf{Bottom Mid}: Shows the radio continuum image, in greyscale, of the same detected galaxy, overlaid with \hi contours.}
    \label{fig:id_9_moment_maps}
\end{figure*}

\textsc{sofia} also provides the zeroth-order and first-order moment maps for all the detected sources. These maps, similar to Figure \ref{fig:moment_maps}, are derived from the masked cube corresponding to each source. In Figure~\ref{fig:id_9_moment_maps} we present both the \hi zeroth-order and first-order moment maps of one of the detected sources (\textsc{sofia} source ID 22) in the MeerKAT \hi data cube. We also present the MeerKAT \hi integrated spectrum for this source. For each of the 16 galaxies with converged kinematic models, we calculate the w$_{50}$ (full width at 50 per cent peak flux) in km\,s$^{-1}$ and these values are presented in Table \ref{tab:properties_table}. 

\subsection{Galaxy masses}

Following \cite{Iorio_2017} we calculate the \hi mass for each of the \hi detected sources using:

\begin{equation}
    \frac{\textit{M}_{\text{H\small{I}}}}{\text{M}_{\odot}} = 23.5 \left( \frac{\delta}{\text{arcsec}} \right)^2 \left( \frac{D}{\text{Mpc}} \right)^2 \sum_{\text{pixels}} \left( \frac{\Sigma_{\text{obs}}(x, y)}{\text{M}_{\odot} \, \text{pc}^{-2}} \right),
    \label{eqn:hi_mass}
\end{equation}

\begin{figure*}
    \centering
    \includegraphics[scale=.45]{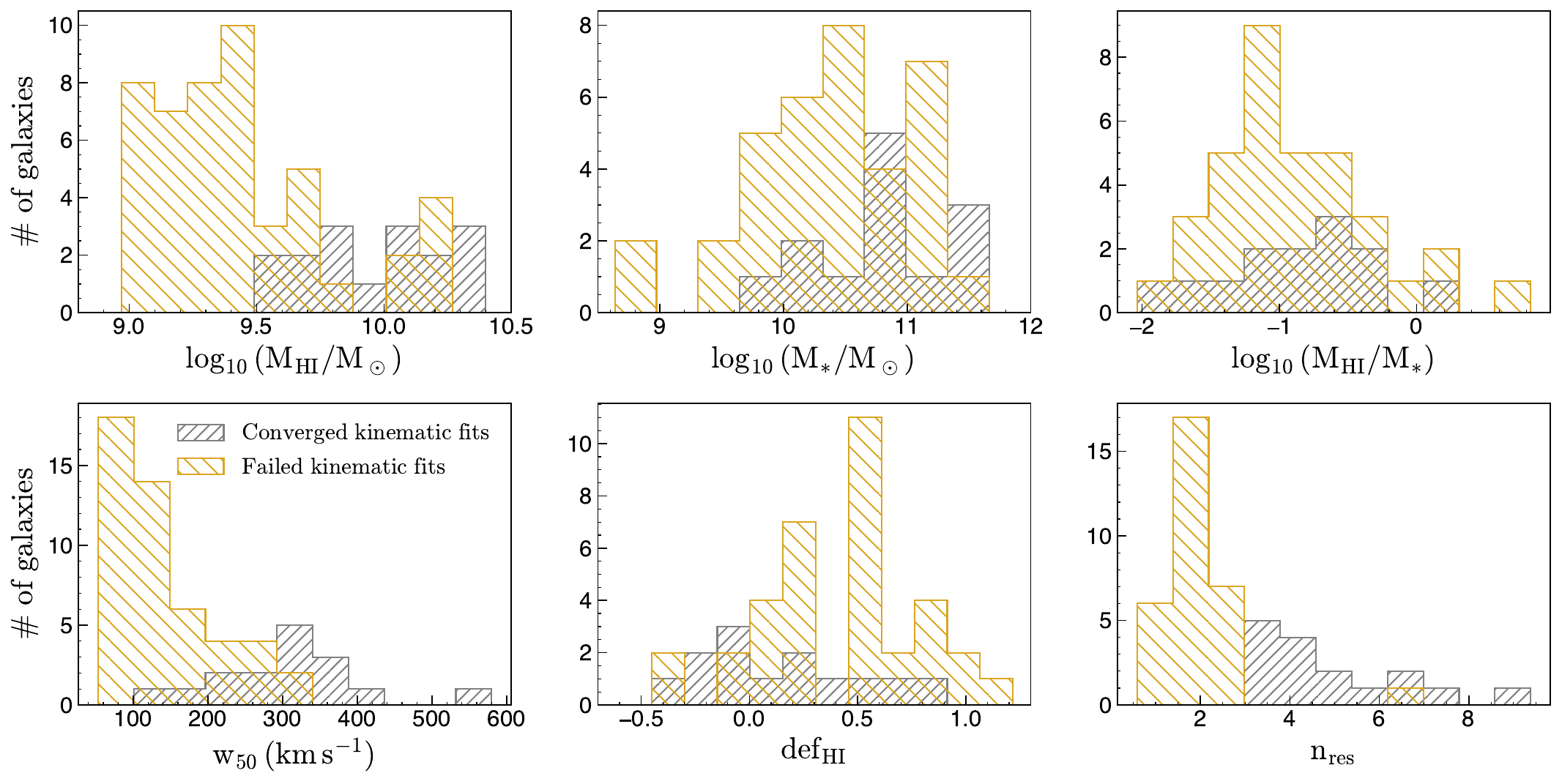}
    \caption{Properties of the MeerKAT \hi detections in the Abell 3408 galaxy cluster. The 16 \hi detections with converged kinematics fits are shown in grey, and the yellow presents the unmodelled detections.}
    \label{fig:detection_properties}
\end{figure*}

\noindent where $\delta$ is size of the pixel, and \textit{D} is the cosmological luminosity distance of the galaxy cluster. Over 50 per cent of the \hi detections have 9.0~$\le$~$\log_{10}$(\textit{M}$_{\rm HI}$/M$_{\odot}$)~$\le$~10, shown in Figure~\ref{fig:detection_properties}. \\

For the  48 \hi detections with counterparts both in the W1 and W2 bands, we calculated the stellar masses (\textit{M}$_{*}$) using Eqn. 1 from \citealt{Cluver_2014}:

\begin{equation}
    \begin{split}
    \indent \rm log_{10} ( \textit{M}_{*}/\textit{L}_{\rm W1}) = -1.97(\textit{W}_{\rm 3.4 \mu m} - \textit{W}_{ \rm 4.6 \mu m}) - 0.03\\
     \textit{L}_{\rm W1}/L_{\odot} = 10^{-0.4(M - M_{\alpha})},
    \end{split}
    \label{eqn:st_mass}
\end{equation}

\noindent where \textit{M} is the absolute magnitude of the source in W1 and \textit{M}$_{\rm \alpha}$~=~3.24.\\ 

\noindent The calculated stellar masses of the 16 galaxies with converged kinematic models are presented in Table \ref{tab:properties_table} and the distribution of \textit{M}$_{*}$ for all the galaxies with WISE counterparts is presented in Figure \ref{fig:detection_properties}. We adopt an indicative uncertainty of $\sim$ 0.1 dex for the calculated stellar masses (\citealt{Ponomareva_2021}), which is consistent with the reported uncertainty of WISE stellar masses $\log_{10}$(\textit{M}$_{*}$/M$_{\odot}$) $>$ 10 (see, \citealt{Jarrett_2023}). Additionally the total dynamical masses presented in Table \ref{tab:properties_table} are calculated using Eqn. 10 from \citealt{Yu_2020}:

\begin{equation}
    \indent \textit{M}_{\rm dyn} = \frac{\textit{V}_{\rm rot}^2\ \textit{R}_{\rm HI} }{G} = 2.33\times10^5 \rm M_{\odot} \left(\frac{\textit{V}_{\rm rot}^2\ \textit{R}_{\rm HI}}{\rm (km\ s^{-1})^2\ kpc }\right),
\end{equation}

\noindent where \textit{G} is the gravitational constant, \textit{V}$_{\rm rot}$ is the \hi rotation curve point, at radius \textit{R}$_{\rm HI}$. \\

\subsection{\hi gas fractions}

We use the \hi and stellar masses to calculate the \hi gas fractions of the individual galaxies, defined as:

\begin{equation}
   f_{\rm g} = M_{\rm HI}/M_{*}.
\end{equation}

\begin{figure}
    \centering
    \includegraphics[scale=.4]{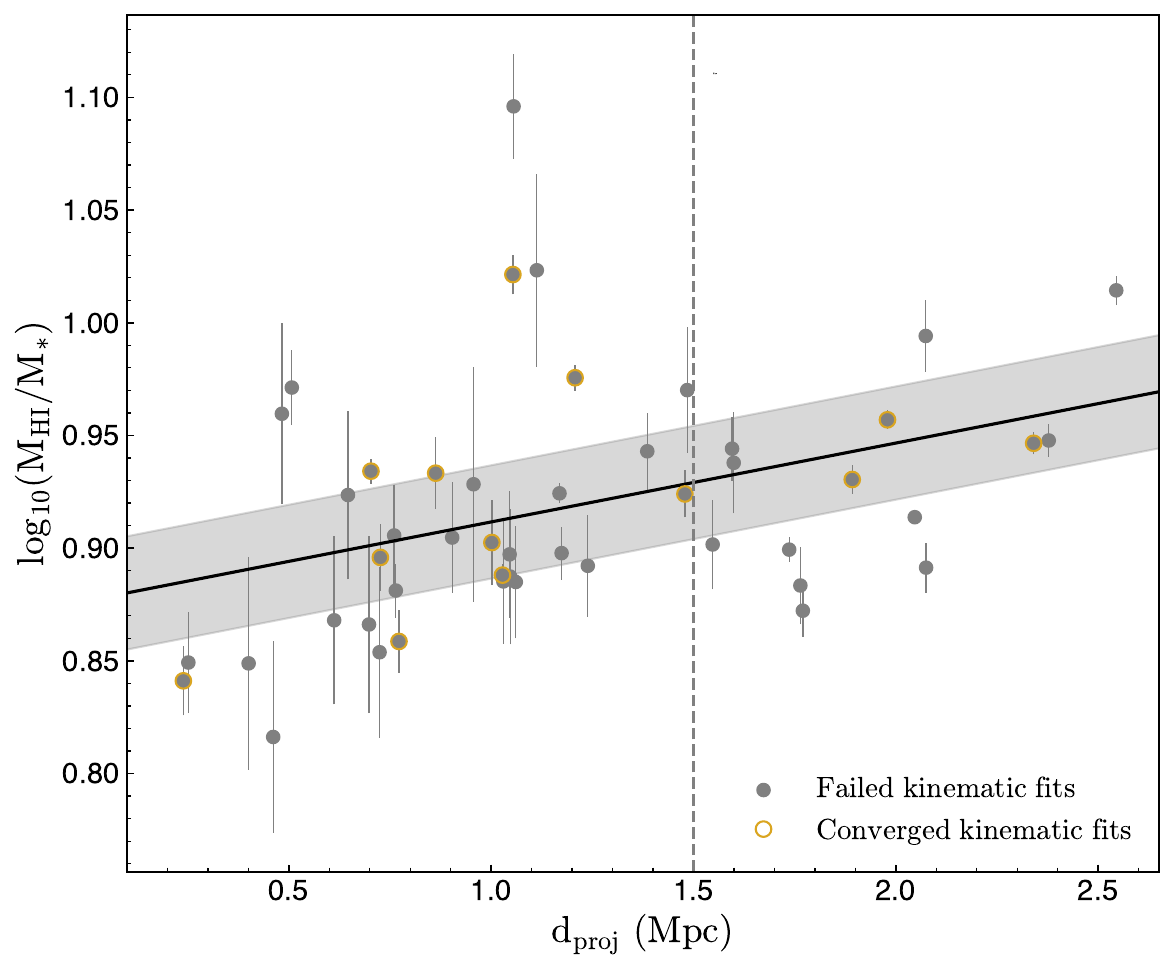}
    \caption{\hi gas fraction of the detections with calculated stellar mass with respect to the galaxies projected cluster-centric distance. The black line presents a linear fit, with 2$\sigma$ scatter shown in grey. The points with yellow annuli mark the galaxies with converged kinematic fits.}
    \label{fig:fg_dproj_fit}
\end{figure}

\noindent The calculated \hi gas fractions are presented in Figure \ref{fig:fg_dproj_fit}, with respect to the projected distance (\textit{d}$_{\rm proj}$) from the cluster centre. We notice a loose trend, such that galaxies in closer proximity (i.e. \textit{d}$_{\rm proj}$~$<$~0.5) to the centre of the cluster have lower gas fractions, and this is seen to increase with the projected distance. A similar result was also found in the studies of \cite{Zabel_2022}, and \cite{moretti_2023}. We perform a linear fit of the following form:

\begin{equation}
    f_{\rm g} = \alpha d_{\rm proj} + \beta,
\end{equation}

\noindent where we achieve $\alpha$ = 0.04, $\beta$ = 0.88 and a scatter of 0.012. The fitted positive gradient suggests that these galaxies are progressively losing their HI content, likely due to some environmental effect.

In addition to this, we calculated the \hi deficiency parameter of the galaxies with WISE counterparts. \hi deficiency is the difference between the \hi observed and expected mass in an isolated galaxy of the same size and morphology (\citealt{Haynes_1984}). It serves as a quantity that indicates the amount of \hi gas in a galaxy. We use Eqn.~\ref{eqn:mhi_exp} to calculate the expected \hi mass, which correlates the observed \textit{M}$_{*}$ of a galaxy (e.g. see \citealt{Catinella_2018}; \citealt{Parkash_2018}).

\begin{equation}
    \log \textit{M}_{\rm HI} = 0.51 (\log \textit{M}_{\rm *} - 10) + 9.71.
    \label{eqn:mhi_exp}
\end{equation}

For the \hi detections with WISE counterparts, the \hi deficiency ranges from -0.5 to $\sim$ 1.2, with more than 50 per cent of the Abell~3408 galaxies falling in the \hi deficiency region (i.e. having $\textit{DEF}_{\rm HI}$~>~0.3, see, \citealt{Denes_2014}; \citealt{Reynolds_2020}). About 39 per cent of the galaxies are found in the normal \hi content regime (i.e. -0.3 < $\textit{DEF}_{\rm HI}$ < 0.3), and only 4 per cent of the galaxies have excess \hi.

\section{\hi kinematics}
\label{sec:4}

We examine how the dense cluster environment may influence the \hi kinematics of individual galaxies detected in Abell 3408. To model the \hi kinematics of the individual galaxies in Abell 3408, we use the 3D tilted ring fitting software, $^{\rm 3D}$$\textsc{Barolo}$ (\citealt{diTeodoro2015sup3D/supGalaxies}). This software fits a tilted-ring model (\citealt{Rogstad_M83}) onto the \hi data cube. A principal feature of the software is that it creates a synthetic data cube based on the input observed cube. Following a ring-by-ring approach, the synthetic data cube is fitted to the \hi data cube. The algorithm iterates through this step minimizing residuals between the data and the model cube. Each of these rings is defined by a set of these kinematic and geometric parameters:

\begin{itemize}
        \item[--] the kinematic centre coordinates in pixels (x$_0$, y$_0$),
        \item[--] \textit{V}$_{\rm rot}$ the gas rotation velocity along circular orbits,
        \item[--] \textit{V}$_{\rm sys}$ the systemic velocity of the galaxy, due to the Hubble flow and its' peculiar velocity,
        \item[--] $\phi$ which is the disc position angle, measured counter-clockwise from the north to the major-axis of the galaxy,
        \item[--] $\textit{i}$ the inclination angle of the gas disc, in the sky plane with respect to a face on circular disc,
        \item[--] \textit{z}$_{0}$ the scale height of the galaxy disc,
        \item[--] \textit{$\Sigma$} face-on \hi column density,
        \item[--] \textit{$\sigma_{\rm HI}$} velocity dispersion.
\end{itemize}

The user can provide initial estimates for each of the parameters, however, $^{\rm 3D}$$\textsc{Barolo}$ can also make initial estimates automatically. These parameters can be left to change for each ring, or can be fixed.

\subsection{Modelling approach}

We develop a pipeline based on \texttt{CANNUBI}\footnote{https://www.filippofraternali.com/cannubi} (\citealt{Roman_Oliveira_2023}) and \texttt{pyBBarolo}\footnote{https://editeodoro.github.io/Bbarolo/pyBBarolo/} to model the \hi kinematics of a large sample of marginally resolved galaxies. This standardized, automated approach is crucial for ensuring both reproducibility and scalability, especially given the growing number of spatially resolved galaxies observed with MeerKAT.\\

\noindent \texttt{CANNUBI}, is a Python-based tool that uses the Markov Chain Monte Carlo (MCMC) sampler \textsc{emcee} (\citealt{Foreman_Mackey_2013}) to robustly determine the geometric parameters of galaxy discs. It estimates the - coordinates of the centre (x$_0$, y$_0$), - the radial extent, - the thickness (Z0), - the position angle (\textit{PA}), and - the inclination angle (\textit{i}) of the galaxy's disc. For more technical details, we refer the reader to \cite{Roman_Oliveira_2023}. \texttt{pyBBarolo} is a recently developed Python wrapper of the main $^{\rm 3D}$$\textsc{Barolo}$ class. \\

\noindent In the first stage of our pipeline, \texttt{pyBBarolo}, is executed in blind mode, meaning that all initial parameter estimates are derived automatically from the input data cube. The kinematic centre coordinates are derived from the zeroth-order moment map (\hi total intensity map) as the flux-weighted average positions of the source. The inclination angle of the \hi\ disc is estimated by fitting a model map, to the observed zeroth-order moment (\hi\ total intensity) map. The position angle is estimated from the \hi velocity field as the line that maximizes the velocity gradient along the line of sight. An initial estimate of the systemic velocity is derived as the central velocity from the global \hi spectrum. The rotation velocity is also derived from the integrated \hi spectrum, following:

\begin{equation}
    \indent V_{\rm rot} = \frac{0.5 \textit{W}_{20}}{\sin \textit{i}},
\end{equation}

\noindent where \textit{W}$_{\rm 20}$ in km\,s$^{-1}$ is the width of the integrated \hi spectrum at 20 per cent of the peak flux and \textit{i} is the inclination angle of the \hi disc.\\

\noindent For all the galaxies, we fixed $\sigma_{\rm HI}$ = 8 $\rm km\,s^{-1}$. We minimize $|\textit{mod} - \textit{obs}|$, where \textit{mod} is the synthetic data cube and \textit{obs} is the MeerKAT \hi datacube, for each detected galaxy. The bulk of rotational information in a galaxy is contained in points closer to the major-axis and thus for all our models we used  $\cos^2 \theta$  weighting function where $\theta$ is the azimuth angle. We use AZIM for normalization of the 3D model and set SNRCUT and GROWTHCUT parameters to 3.5 and 3, respectively. The outputs of this first blind run are then used to generate parameter files, for each galaxy.\\ 

\noindent In the second step of the pipeline, the parameter files generated in the first step are used as input to run \texttt{CANNUBI} for each galaxy. The routine fits the central coordinates (x$_0$, y$_0$), inclination, position angle, and ring separation (RADSEP) using 30 MCMC walkers over 1000 iterations. Key \texttt{CANNUBI} configuration parameters are set as follows: the masking method is set to SMOOTH$\&$SEARCH, the noise level is computed dynamically for each data cube, the FACTOR is fixed at 1.5, BLANKCUT at 3, SNRCUT at 3.5, and GROWTHCUT at 2. The resulting fitted parameters are saved in an output file specific to each galaxy.\\

\begin{figure*}
    \centering
    \includegraphics[scale=.5]{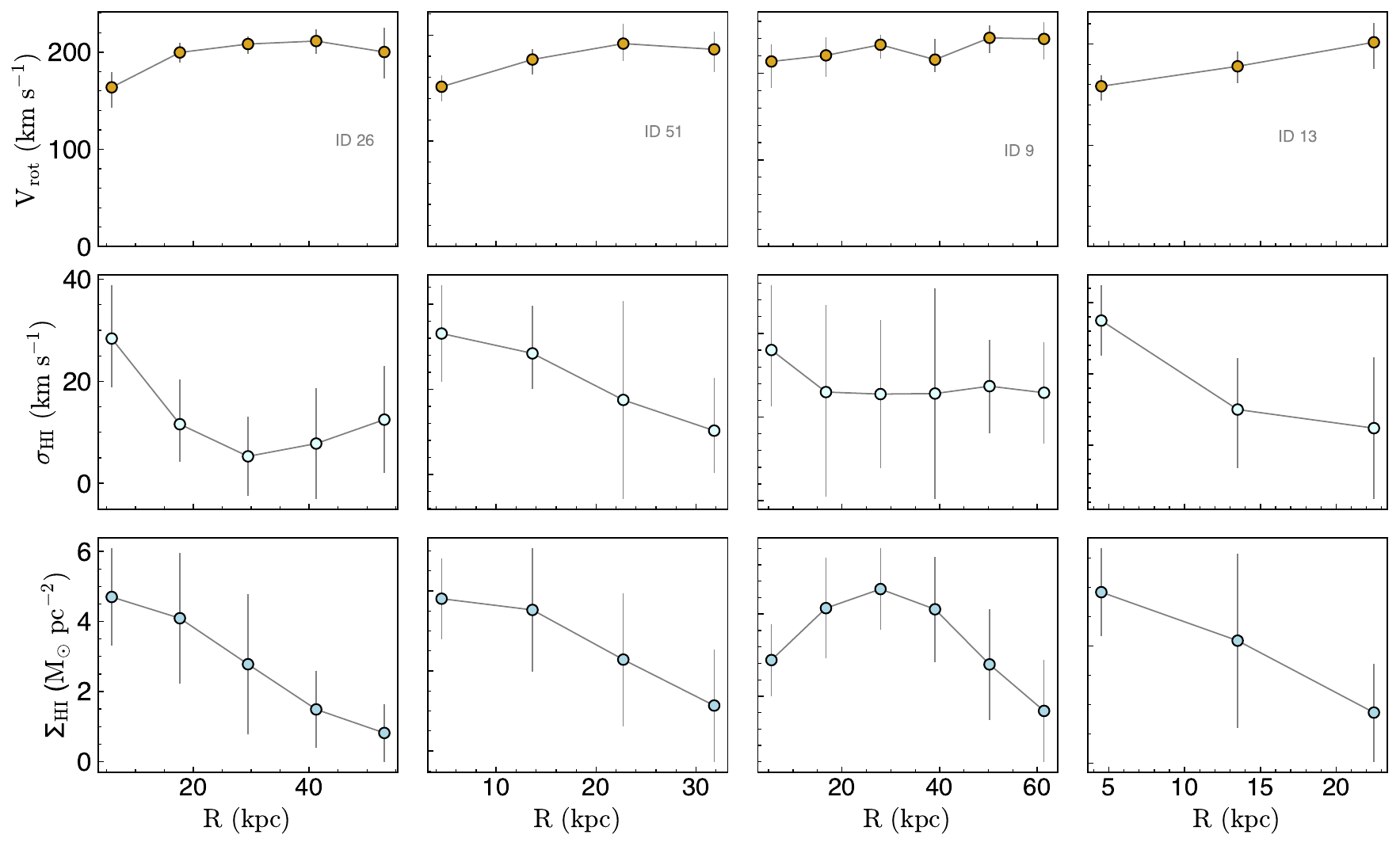}
    \caption{Data for four galaxies representative of the sample of modelled galaxies. The top row presents the \hi rotation curves, \hi velocity dispersion profile is on the middle row, and the radial \hi surface density profiles in the bottom panel.}
    \label{fig:vrot_vdisp}
\end{figure*}

\noindent In the final step, we use the fitted parameters from \texttt{CANNUBI} to model the \hi rotation curves of the galaxies with \texttt{pyBBarolo}. This step is configured similarly to the initial blind run, except that the inclination, position angle, and ring width (RADSEP) are fixed to the values obtained from \texttt{CANNUBI}. We adopt an initial value of 8 km\,s$^{-1}$ for the velocity dispersion (V$_{\rm disp}$), which is fitted simultaneously with the rotation curve (Figure \ref{fig:vrot_vdisp}).\\

\noindent Among the 64 \hi detections, only 19 galaxies span at least three independent resolution elements. We exclude galaxies with fitted inclination angles below 30$^\circ$. After applying this criterion, 16 galaxies remain, for which we successfully construct \hi kinematic models.

\subsection{Results}

\begin{figure*}
    \centering
    \includegraphics[scale=.3]{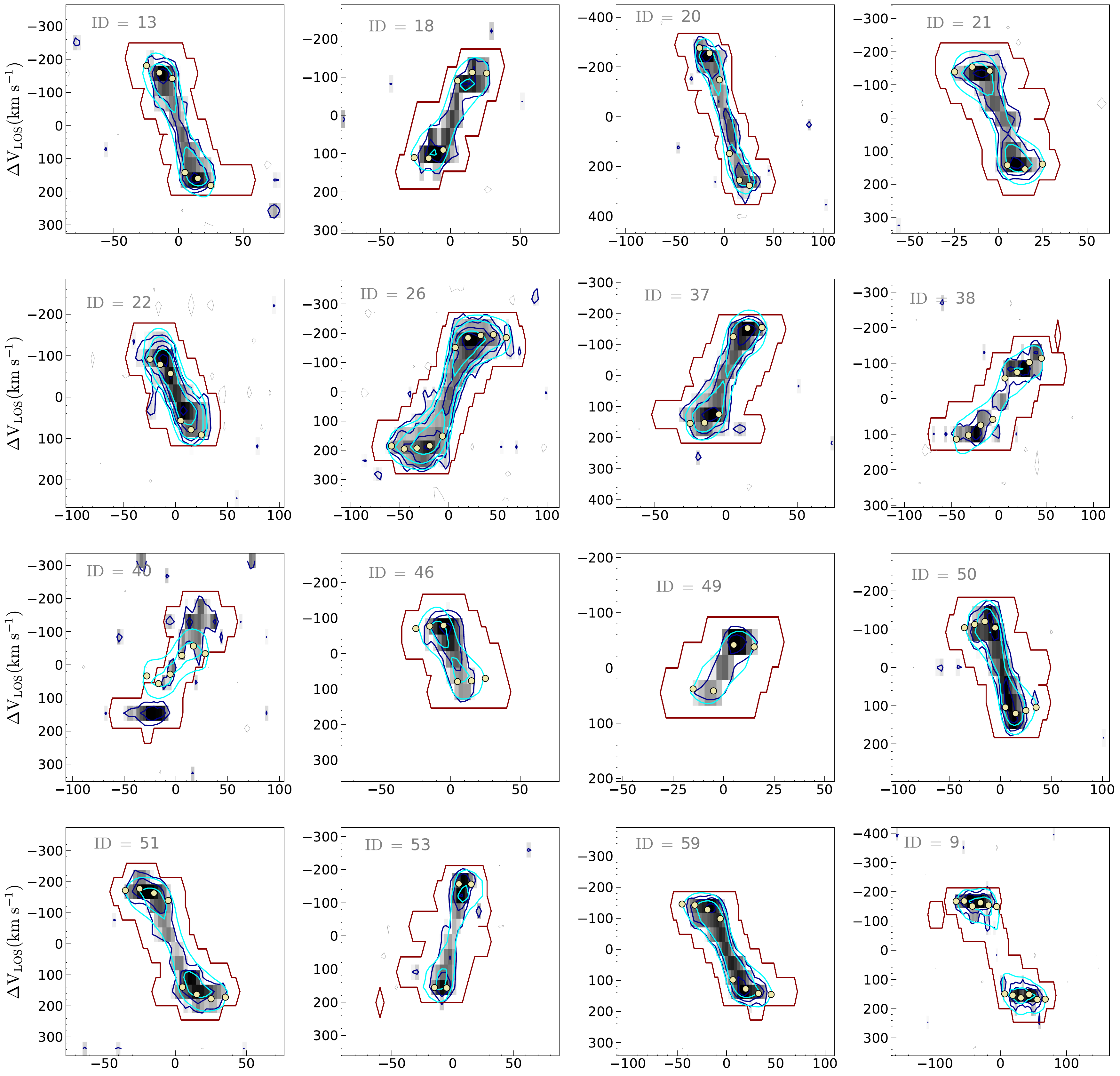}
    \caption{Rotation curves (yellow dots) of the 16 modelled galaxies overlaid on the \hi major axis position-velocity slices (greyscale with dark-blue contours). The cyan contours present the major-axis position-velocity slice extracted from the $^{\rm 3D}$\textsc{Barolo} model, with the mask shown with maroon contours. The contour levels begin at \textit{S}$_{\rm HI}$ = 188.4 $\mu$Jy\ beam$^{\rm -1}$ and increase in multiples of 2.}
    \label{fig:pv_slices}
\end{figure*}

\begin{figure*}
    \centering
    \includegraphics[scale=.87]{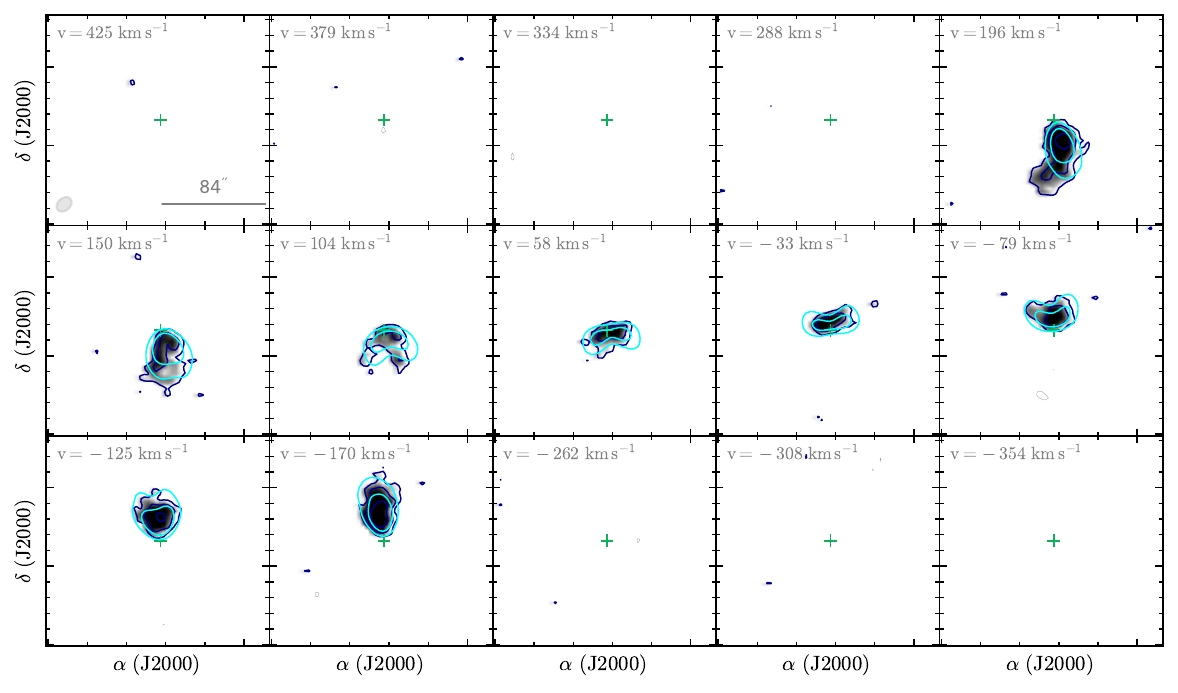}
    \caption{Galaxy ID 26 channel by channel comparison of the data, in greyscale with the dark-blue contours, and the $^{\rm 3D}$$\textsc{Barolo}$ model presented with cyancontours. The green cross marks the $^{\rm 3D}$$\textsc{Barolo}$ model centre coordinates. The MeerKAT synthesized beam (13.1 $\times$ 11.7 arcsec) is displayed in the top left panel, with the grey horizontal bar representing the physical scale.}
    
    \label{fig:channel_by_channel_comp}
\end{figure*}

Figure \ref{fig:pv_slices} presents a comparison of the data with best-fit kinematic models. As seen in Figure~\ref{fig:pv_slices}, the best-fit models reproduce reasonably well with the emission for each of the modelled galaxies. In Figure~\ref{fig:channel_by_channel_comp}, we present channel-by-channel comparison of the data and the $^{\rm 3D}$BBarolo kinematic model for a single galaxy. The model effectively reproduces the observed emission. Table~\ref{tab:properties_table} presents properties of the 16 galaxies with converged \hi kinematic models. Two galaxies (IDs: 40 and 46)  were excluded from our sample of modelled galaxies for subsequent analysis. These galaxies exhibit kinematic models with unphysical rotation curves, based on visual inspection. Furthermore, the \hi\ moment maps of both galaxies display disturbed morphologies, supporting their exclusion. The derived rotation curve data outer points for all the modelled galaxies are presented in Table~\ref{tab:circ_velocities}.\\

\subsection{Rotation curve shapes}

Given that we have generated the first rotation curves for these cluster galaxies we now study how the distribution of luminous matter scales with the rotation curve shapes. The distribution of luminous matter is known to impact the shape of the rotation curves for all galaxy types.  More luminous galaxies are known to have steep rising rotation curves in the inner parts followed by a flatter gradient at large radii and less luminous galaxies (e.g. dwarf galaxies) have been reported to have rotation curves with shallower slopes in the inner regions, and rising to the last measured point (e.g. \citealt{Swaters_2012}).

In our work, we look at the \textit{M}$_{*}$ of the individual galaxies in an attempt to investigate the variation of the rotation curves at different \textit{M}$_{*}$ bins.\\

\begin{figure*}
    \centering
    \includegraphics[scale=.55]{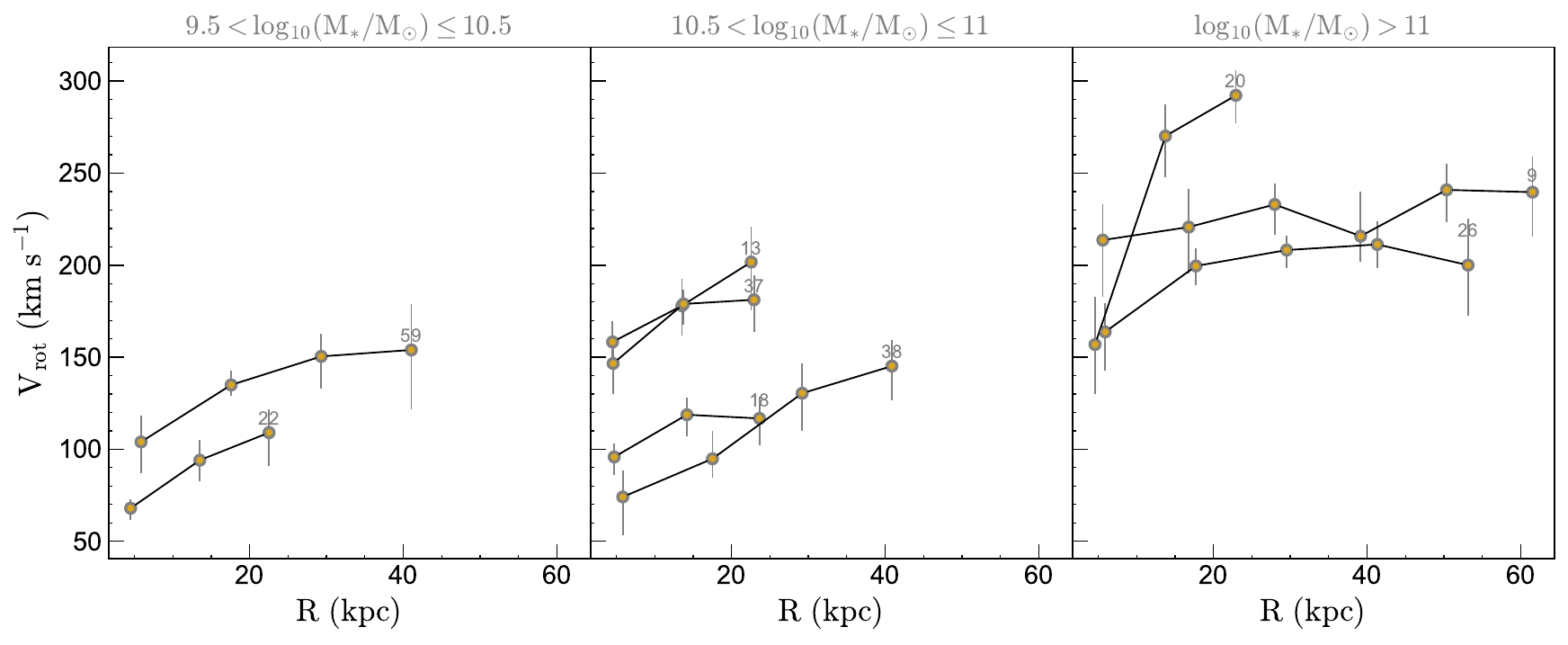}
    \caption{Rotation curves of 9 galaxies with calculated stellar masses, at different stellar mass bins.}
    \label{fig:vrot_mstar}
\end{figure*}

In Figure~\ref{fig:vrot_mstar}, we present the rotation curves of a subset of 9 galaxies with calculated stellar masses (\textit{M}$_{*}$). We study the shapes of these 9 rotation curves in three stellar mass bins, as shown in Figure~\ref{fig:vrot_mstar}. For galaxies in the stellar mass range (9.5~$<$~$\log_{10}$(\textit{M}$_{*}$/M$_{\odot}$)~$<$~10.5), the rotation curves exhibit a gradual rise in the inner regions and continue to increase out to the last measured point. This behaviour is consistent with the study of \cite{Swaters_2012} and suggests a dominant dark matter component throughout the galaxy. The dark matter halo likely provides the majority of the gravitational potential, resulting in a shallow inner slope and a steadily rising curve at larger radii, where baryonic matter plays a relatively minor role.  \\

For intermediate-mass galaxies (10.5~$<$~$\log_{10}$(\textit{M}$_{*}$/M$_{\odot}$)~$<$~11), the rotation curves also tend to rise throughout the observed radial range. While two galaxies in this mass range appear to show some flattening in the outer regions, this is based on only three rotation curve points per galaxy, with the final two points showing a flat gradient. As such, these cases provide limited evidence for a clear transition to dark matter dominance at large radii. Overall, the general shape of the rotation curves in this mass range does not differ significantly from that of the lower-mass systems in this cluster, possibly indicating that dark matter continues to play a significant role even in the inner regions, or that the stellar and gas contributions are more modest than previously inferred.\\
 
For high-mass galaxies ($\log_{10}$(\textit{M}$_{*}$/M$_{\odot}$)~$>$~11), the rotation curves generally flatten at larger radii, consistent with dark matter dominating the outer kinematics. Although a steep central rise could suggest a strong central concentration of baryons, in our sample only one galaxy shows a pronounced inner velocity increase.  Furthermore, the coarse velocity resolution (46 km\,s$^{-1}$) may lead to residual beam smearing, which can smooth out steep velocity gradients in the innermost regions of the rotation curves. \\

\section{Baryonic tully-fisher relation}
\label{sec:5}

We study the Baryonic Tully-Fisher Relation (bTFr) for galaxies with converged kinematic models, examining the \textit{V}$_{\rm out}$ definition. Following the approach of \cite{Ponomareva_2021}, we define \textit{V}$_{\rm out}$ as the rotation velocity at the outermost \hi radius. The uncertainties on \textit{V}$_{\rm out}$ are derived from the uncertainties obtained in the kinematic modelling.
The baryonic mass $M_{\rm bar}$ is calculated as:

\begin{equation}
    M_{\rm bar} = 1.33M_{\rm HI} + M_{*},
    \label{eqn:m_bar}
\end{equation}

\noindent where \textit{M}$_{\rm HI}$ is the \hi mass and \textit{M}$_{*}$ is the stellar mass, calculated with Eqns. \ref{eqn:hi_mass} and \ref{eqn:st_mass} respectively. To account for the contribution of Helium we use the factor 1.33 and neglect the molecular gas since it generally contributes less than 10 per cent of \textit{M}$_{\rm bar}$ (\citealt{McGaugh_2015}). We calculate the uncertainties on \textit{M}$_{\rm bar}$ following Eqn. 5 of \cite{Lelli_2016}. We perform a linear fit of the form:

\begin{equation}
    \log(M_{\rm bar}) = \alpha \log(x) + \beta,
\end{equation}

\noindent where \textit{M}$_{\rm bar}$ is the baryonic mass as defined by Eqn. \ref{eqn:m_bar}, \textit{x} is the velocity definition, \textit{V}$_{\rm out}$, $\alpha$ is the slope, and $\beta$ is the zero-point. \\                                                                 

\begin{table}
	\begin{center}
         \caption{Rotation curve data of the 14 galaxies with converged kinematic fits.}
		\begin{tabular}{|c|c|c|c|c|c|}
			\hline
			ID & $r_{\rm out}$ & $v_{\rm out}$ & $i$ & $\rm PA $ & $v_{\rm sys}$ \\
            -- & kpc & km\ s$^{-1}$ &  $^{\circ}$ & $^{\circ}$ & km\,s$^{-1}$ \\
            \small (1) & (2) & (3) & (4) & (5) & (6) \\
			\hline
			9 & 61.54 & 239.71 & 44.30 & 304.62 & 707.54 \\
			13 & 22.56 & 201.72 & 63.64 & 353.49 & 559.76 \\
			18 & 23.64 & 116.72 & 70.91 & 109.92 & 162.87 \\
			20 & 22.92 & 292.29 & 71.03 & 330.24 & 369.68 \\
			21 & 22.55 & 174.48 & 52.93 & 266.07 & 491.03 \\
			22 & 22.53 & 109.02 & 56.87 & 228.01 & 388.59 \\
			26 & 53.15 & 200.04 & 67.73 & 190.22 & -384.04 \\
			37 & 22.92 & 181.22 & 58.34 & 86.29 & -45.58 \\
			38 & 40.88 & 145.17 & 51.71 & 71.31 & -110.76 \\
			49 & 13.75 & 63.92 & 36.54 & 87.51 & -653.93 \\
			50 & 31.90 & 128.91 & 53.82 & 41.55 & -654.68 \\
			51 & 31.90 & 186.55 & 67.16 & 357.15 & -440.94 \\
			53 & 13.67 & 221.20 & 44.82 & 210.10 & -764.02 \\
			59 & 41.05 & 153.96 & 70.92 & 261.44 & -791.02 \\
			\hline
		\end{tabular}
            \label{tab:circ_velocities}

	\end{center}
       
\end{table}

\begin{table}
    \caption{The fitted statistical properties of the bTFr.}
    \resizebox{0.7\columnwidth}{!}{
	\begin{tabular}{lc} 
		\hline
		\hline\\
		Slope & 3.66$^{+0.32}_{-0.28}$\\
        \\[-0.5em] 
		Zero point  & 1.73$^{+0.82}_{-0.59}$\\
        \\[-0.5em] 
		Intrinsic scatter ($\sigma_{\perp}$) & 0.08$^{+0.01}_{-0.01}$\\
    \hline
	\end{tabular}}
 	\label{tab:bTFr_tab}
\end{table}

\begin{figure*}
    \centering
    \includegraphics[scale=.9]{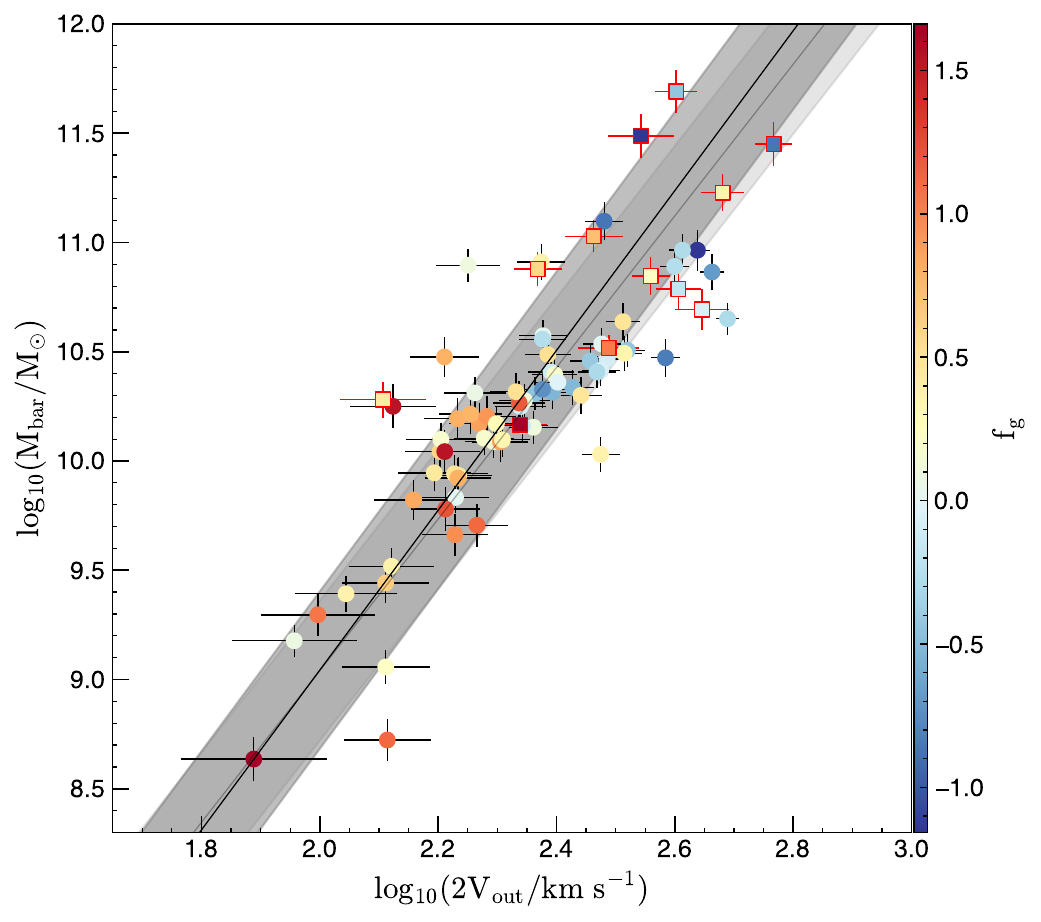}    
    \caption{Baryonic Tully-Fisher Relation for the outermost rotation curve (\textit{V}$_{\rm out}$) velocity definition. The points are coloured by \textit{f}$_{\rm g}$, and the black line shows linear fit to the data. The dark gray shaded area shows the 1-$\sigma$ uncertainty of the fit. The data points from the MIGHTEE bTFr are shown with circles and squares for the Abell 3408 selected galaxies with converged kinematic fits. The grey line with the light grey shaded area is the linear fit and 1-$\sigma$ uncertainty of the MIGHTEE bTFr.}
    \label{fig:bTFr}
\end{figure*}

\begin{figure}
    \centering
    \includegraphics[scale=.36]{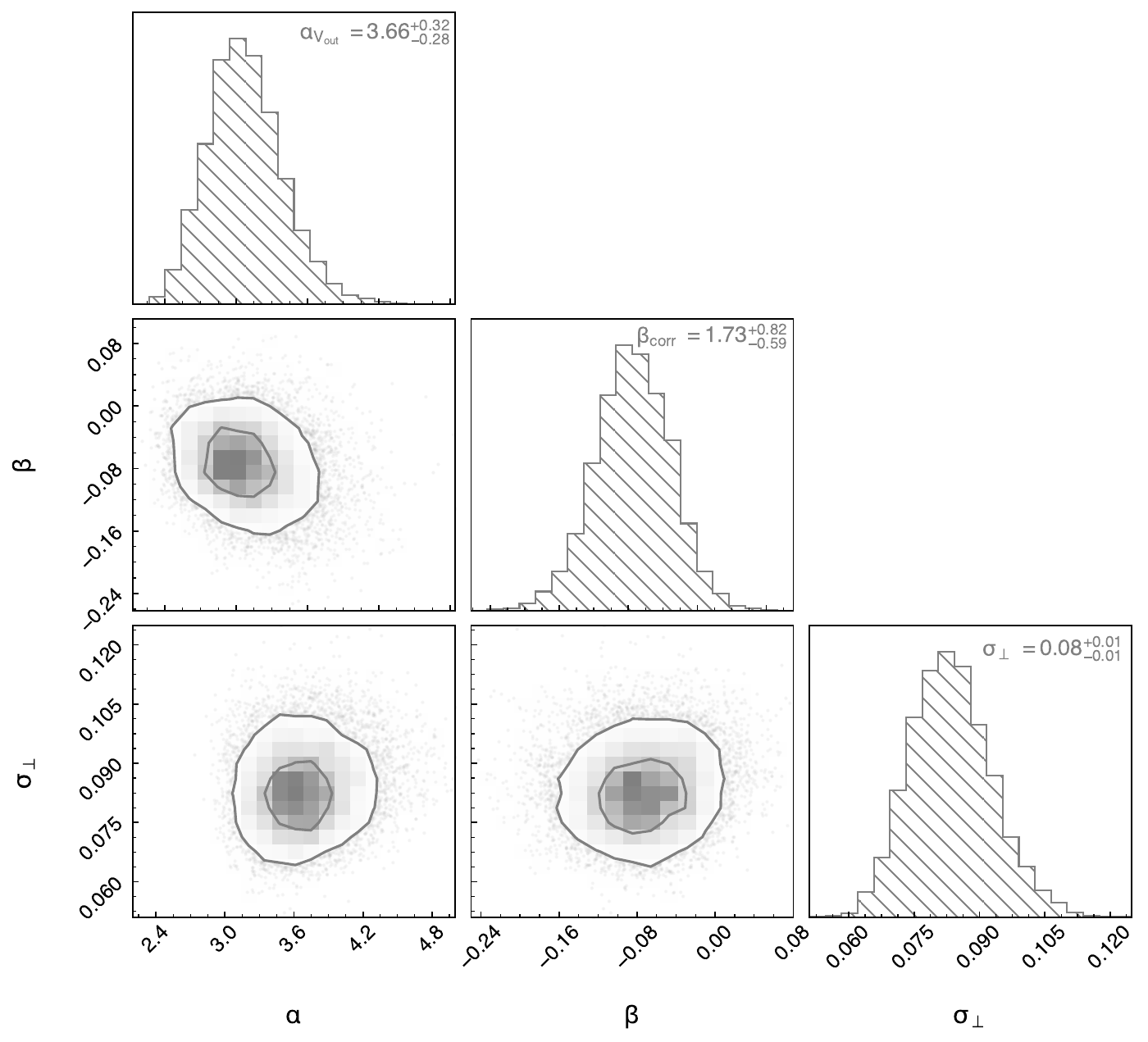} 
    \caption{The posterior distributions of the slope ($\alpha$), zero point ($\beta$) and the orthogonal scatter ($\sigma_{\perp}$) for the bTFR based on \textit{V}$_{\rm out}$. The contour levels are at 68 and 95 per cent confidence intervals.}
    \label{fig:post_dist}
\end{figure}

In this paper we extend our sample of galaxies with that of \cite{Ponomareva_2021}, resulting in a total of 79 galaxies, spanning a redshift range 0.006 $\lesssim$ \textit{z} $\lesssim$ 0.081. We leverage the methodology described by \cite{Lelli_2019}. We employ a well-established Monte Carlo Markov Chain (MCMC) technique implemented in \textsc{emcee} (\citealt{Foreman_Mackey_2013}) to sample the posterior probability distributions of crucial statistical parameters: $\alpha$, $\beta$, and $\sigma_{\perp}$. We initialise our MCMC simulation with 50 walkers. Each walker explores the parameter space for three parameters: $\alpha$, $\beta$, and $\sigma_{\perp}$. To start, the walkers are positioned randomly within pre-defined ranges that reflect our prior knowledge about these parameters. The $\alpha$ can range from 3.0 to 5.0, the $\beta$ from 1.0 to 5.0, and the $\sigma_{\perp}$ from 0.01 to 0.25, all of which have uniform prior probabilities within these limits. The sampler is then run for 1000 iterations, followed by another 1000 iterations to ensure thorough exploration. Figure~\ref{fig:post_dist} presents the posterior distributions of the fitted parameters for the bTFr. The best-fit values of the parameters, for both velocity definitions, are presented in Table~\ref{tab:bTFr_tab}. \\

Figure \ref{fig:bTFr} shows the bTFr for the combined sample i.e. 79 galaxies, for the \textit{V}$_{\rm out}$ velocity definition. We find that the properties of the relation using this velocity definition are consistent with those achieved in \cite{Ponomareva_2021}. Notably, \hi detections from the Abell~3408 galaxy cluster, despite being at relatively low redshifts ($\langle$~\textit{z}~$\rangle$~$\sim$~0.042) extend the bTFr in both mass and velocity space. Several of these Abell 3408 \hi detections are falling within the 1$\sigma$ scatter of the relation, with few seen below and above the relation.\\

\section{Summary and Conclusions}
\label{sec:6}

In this paper, we use MeerKAT \textit{L}-band observations to study the kinematics of individual galaxies in the Abell 3408 galaxy cluster at \textit{z} $\sim$ 0.042, \textit{D} $\sim$ 187 Mpc. Using \textsc{sofia} (\citealt{Serra_2015}) a total of 64 galaxies are detected in \hi in this X-ray luminous cluster. The main results of this work are summarised as follows:

\begin{itemize}

    \item We study the impact of the dense cluster environment on the global properties of the galaxies. We measure that galaxies closer to the cluster centre exhibit higher degree of \hi deficiency compared to galaxies seen over a projected distance of \textit{d}$_{\rm proj}$~$\sim$~0.5 Mpc. We perform a linear fit to this observed trend and find a relation of \textit{f}$_{g}$~=~0.04\textit{d}$_{\rm proj}$ + 0.88, with a scatter of 0.012.\\

    \item We present rotation curves of cluster galaxies generated in a uniform way. To study the \hi kinematics of this sample of galaxies, we develop a semi-automated pipeline, that requires little to no manually chosen priors about the galaxy's kinematic parameters and human intervention. Despite the coarse spatial and velocity resolution (46 km\,s$^{-1}$) of the MeerKAT \hi datacubes at hand, our pipeline is able to model the rotation and dispersion curves for several galaxies with the produced \hi kinematic models consistent with the data.\\

    \item Low-mass galaxies (log$_{10}$(\textit{M}$_{*}$/M$_{\odot}$)~$<$~10) exhibit rotation curves that rise gradually from the inner regions and continue to increase out to the last measured point, consistent with dark matter dominance throughout their radial extent. Intermediate-mass galaxies (10~$<$~log$_{10}$(\textit{M}$_{*}$/M$_{\odot}$)~$<$~11) show similar rising rotation curves, with only limited evidence for flattening in the outer regions, seen in two cases with sparse sampling (three data points), making it difficult to draw firm conclusions about baryonic versus dark-matter dominance. High-mass galaxies (log$_{10}$(\textit{M}$_{*}$/M$_{\odot}$)~$>$~11) display a sharp rise in the central regions, although this feature is seen in only one galaxy and may be affected by beam smearing, limiting strong claims about central baryonic concentration.\\

    \item Based on the (\textit{V}$_{\rm out}$ velocity definitions, we study the Baryonic Tully-Fisher Relation (bTFr) in Abell 3408. We extend our sample of galaxies with that of the MIGHTEE Survey Early Science data (\citealt{Ponomareva_2021}), resulting in 79 galaxies in total, spanning a broad range of environments. For the bTFr based on \textit{V}$_{\rm out}$ we achieve a tight relation $\sigma_{\perp}$ = 0.08$^{+0.01}_{-0.01}$ which is consistent within errors to the one found in \cite{Ponomareva_2021}, similarly for the slope and the zero-point. The MeerKAT \hi detections from the Abell 3408 galaxy extend the MIGHTEE bTFr in both mass and velocity spaces. \\

    \item The MeerKAT telescope continuously probes vast cosmological volumes, generating enormous extragalactic datasets. With the upcoming MeerKAT+ and SKA-mid expansions, these datasets will increase dramatically. Therefore, automated tools for detailed analysis of these datasets are essential.
    
\end{itemize}

\section*{Acknowledgements}
We thank the anonymous referee for their useful and in-depth comments 
that have greatly improved the quality of this manuscript. The authors would like to thank Anastasia Ponomareva for providing (through private communication) the MIGHTEE bTFr data and for the useful discussions on this section. We would also like to thank Enrico Di Teodoro and Filippo Fraternali for their useful inputs on \textsc{pyBBarolo} and \textsc{CANNUBI}. We would like to thank the staff of the South African Radio Astronomy Observatory (SARAO\footnote{www.sarao.ac.za}) who made these observations possible. XN and RPD acknowledge funding from the South African Radio Astronomy Observatory (SARAO), which is a facility of the National Research Foundation (NRF), an agency of the Department of Science and Innovation (DSI). RPD’s research is funded by the South African Research Chairs Initiative of the DSI/NRF.
The MeerKAT telescope is operated by the South African Radio Astronomy Observatory (SARAO), which is a facility of the National Research Foundation, an agency of the Department of Science and Innovation. The authors also acknowledge the use of the ilifu cloud computing facility – \href{https://www.ilifu.ac.za/}{www.ilifu.ac.za}, a partnership between the University of Cape Town, the University of the Western Cape, the
University of Stellenbosch, Sol Plaatje University, the Cape
Peninsula University of Technology and the South African
Radio Astronomy Observatory. The Ilifu facility is supported
by contributions from the Inter-University Institute for Data
Intensive Astronomy (IDIA – a partnership between the University of Cape Town, the University of Pretoria, the University of the Western Cape and the South African Radio Astronomy Observatory), the Computational Biology division at UCT, and the Data Intensive Research Initiative of South
Africa (DIRISA). This work has made use of the Cube Analysis and Rendering Tool for Astronomy \citep[CARTA,][]{Comrie_2021}. This research made use of Astropy\footnote{http://www.astropy.org}, a community-developed core Python package for Astronomy. This research makes use of the semi-automated MeerKAT data calibration pipeline, \textsc{Oxkat}\footnote{https://github.com/IanHeywood/oxkat}. \\

\section*{Data Availability}

The MeerKAT reduced cubes and derived data products used in this work are available upon request from corresponding authors.

\bibliographystyle{mnras}
\bibliography{example}

\appendix

\section{Appendix A: MeerKAT moment 0 maps of all the detections}
\label{sec:appendix_a}

We present the MeerKAT \hi moment 0 maps for all the 64 detections of the Abell 3408 galaxy cluster. The order of the moment 0 is by increasing stellar masses. 

\begin{figure*}
    \centering
    \includegraphics[width=.89\textwidth, height=\textwidth]{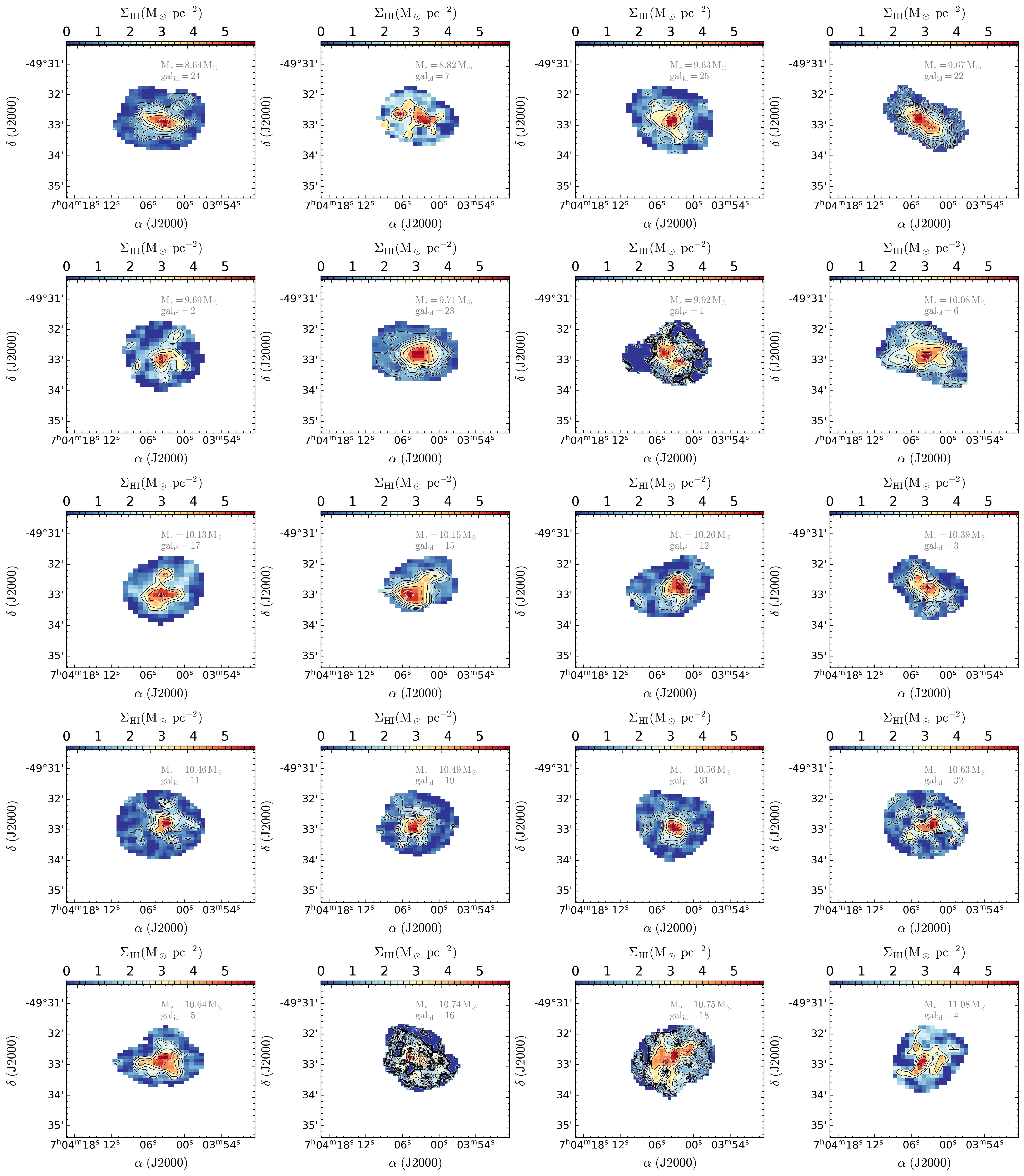}
    \caption{MeerKAT \hi moment 0 maps of all the 64 detections ordered by increasing stellar masses.}
    \label{fig:maps}
\end{figure*}

\begin{figure*}
    \ContinuedFloat
    \centering
    \includegraphics[width=.89\textwidth, height=\textwidth]{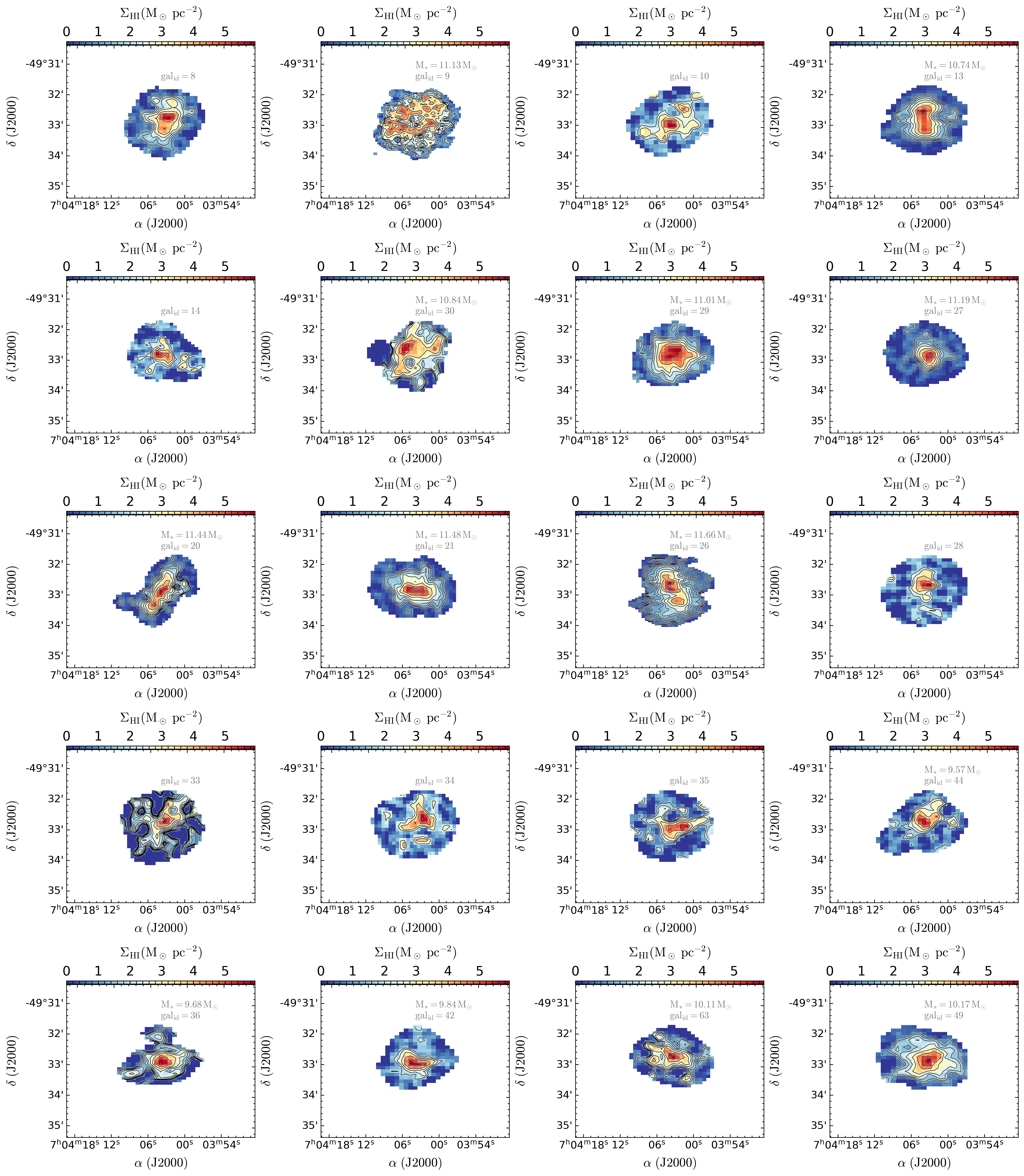}
    \caption{Continued}
    \label{fig:maps}
\end{figure*}

 \begin{figure*}
    \ContinuedFloat
    \centering
    \includegraphics[width=.85\textwidth, height=\textwidth]{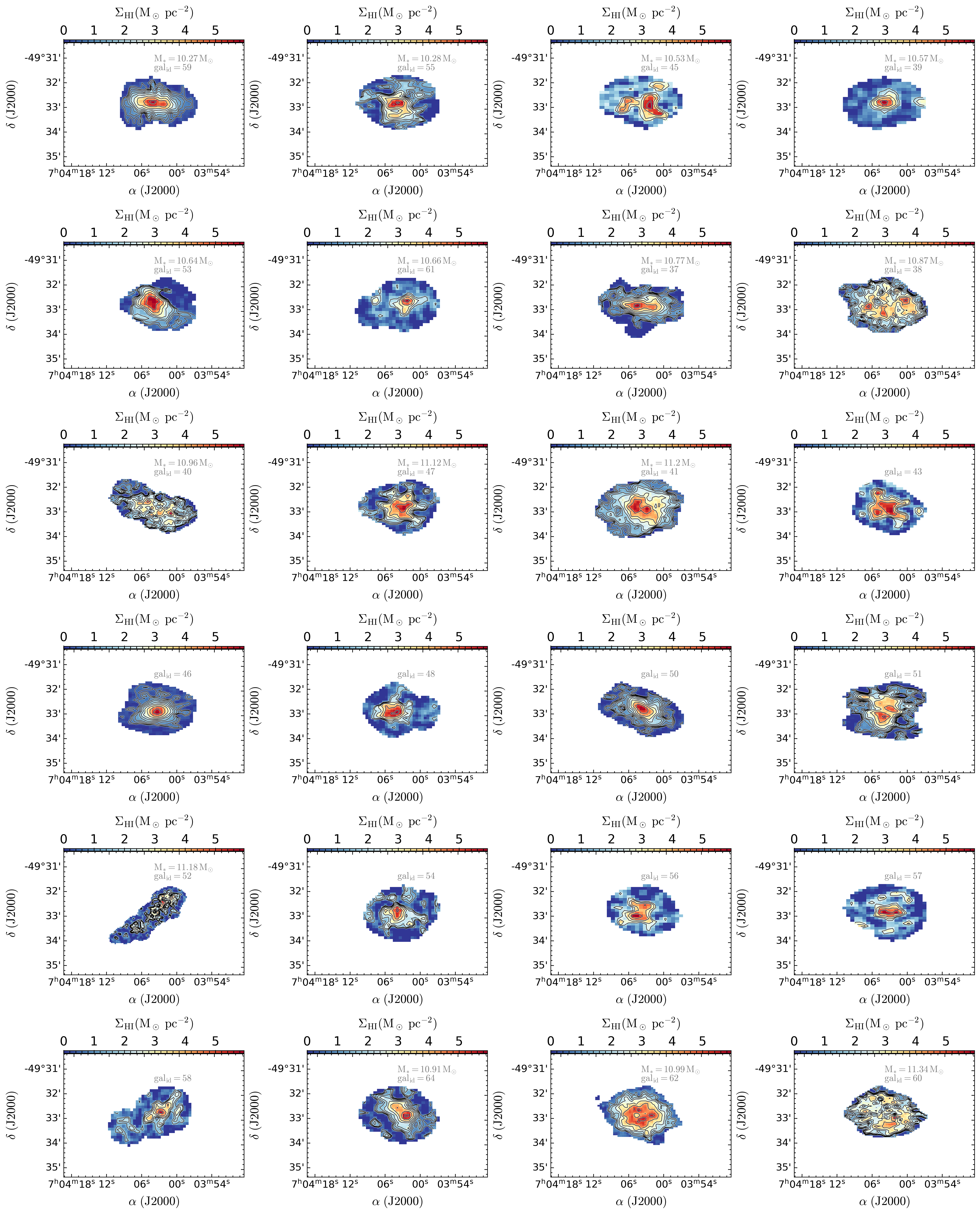}
    \caption{Continued}
    \label{fig:maps}
\end{figure*}

\bsp	
\label{lastpage}
\end{document}